\documentclass[twoside,twocolumn,9pt]{article}
\usepackage{extsizes}
\usepackage[super,sort&compress,comma]{natbib} 
\usepackage[version=3]{mhchem}
\usepackage[left=1.5cm, right=1.5cm, top=1.785cm, bottom=2.0cm]{geometry}
\usepackage{balance}
\usepackage{mathptmx}
\usepackage{sectsty}
\usepackage{graphicx} 
\usepackage{lastpage}
\usepackage[format=plain,justification=justified,singlelinecheck=false,font={stretch=1.125,small,sf},labelfont=bf,labelsep=space]{caption}
\usepackage{float}
\usepackage{fancyhdr}
\usepackage{fnpos}
\usepackage{breqn}
\usepackage{comment}
\usepackage[english]{babel}
\addto{\captionsenglish}{%
  
}
\usepackage{array}
\usepackage{droidsans}
\usepackage{charter}
\usepackage[T1]{fontenc}
\usepackage[usenames,dvipsnames]{xcolor}
\usepackage{setspace}
\usepackage[compact]{titlesec}
\usepackage{hyperref}
\usepackage{mathtools}
\usepackage{amsmath}
\usepackage{xcolor}
\hypersetup{
    colorlinks,
    linkcolor={red!50!black},
    citecolor={blue!50!black},
    urlcolor={blue!80!black}
}

\usepackage{epstopdf}

\definecolor{cream}{RGB}{222,217,201}

\begin{document}

\pagestyle{fancy}
\thispagestyle{plain}
\fancypagestyle{plain}{
\renewcommand{\headrulewidth}{0pt}
}

\makeFNbottom
\makeatletter
\renewcommand\LARGE{\@setfontsize\LARGE{15pt}{17}}
\renewcommand\Large{\@setfontsize\Large{12pt}{14}}
\renewcommand\large{\@setfontsize\large{10pt}{12}}
\renewcommand\footnotesize{\@setfontsize\footnotesize{7pt}{10}}
\makeatother

\renewcommand{\thefootnote}{\fnsymbol{footnote}}
\renewcommand\footnoterule{\vspace*{1pt}%
\color{cream}\hrule width 3.5in height 0.4pt \color{black}\vspace*{5pt}}
\setcounter{secnumdepth}{5}

\makeatletter 
\renewcommand\@biblabel[1]{#1}            
\renewcommand\@makefntext[1]%
{\noindent\makebox[0pt][r]{\@thefnmark\,}#1}
\makeatother 
\renewcommand{\figurename}{\small{Figure}~}
\sectionfont{\sffamily\Large}
\subsectionfont{\normalsize}
\subsubsectionfont{\bf}
\setstretch{1.125} 
\setlength{\skip\footins}{0.8cm}
\setlength{\footnotesep}{0.25cm}
\setlength{\jot}{10pt}
\titlespacing*{\section}{0pt}{4pt}{4pt}
\titlespacing*{\subsection}{0pt}{15pt}{1pt}

\fancyfoot{}
\fancyfoot[LO,RE]{\vspace{-7.1pt}\includegraphics[height=9pt]{head_foot/LF}}
\fancyfoot[CO]{\vspace{-7.1pt}\hspace{11.9cm}\includegraphics{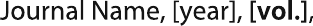}}
\fancyfoot[CE]{\vspace{-7.2pt}\hspace{-13.2cm}\includegraphics{head_foot/RF}}
\fancyfoot[RO]{\footnotesize{\sffamily{1--\pageref{LastPage} ~\textbar  \hspace{2pt}\thepage}}}
\fancyfoot[LE]{\footnotesize{\sffamily{\thepage~\textbar\hspace{4.65cm} 1--\pageref{LastPage}}}}
\fancyhead{}
\renewcommand{\headrulewidth}{0pt} 
\renewcommand{\footrulewidth}{0pt}
\setlength{\arrayrulewidth}{1pt}
\setlength{\columnsep}{6.5mm}
\setlength\bibsep{1pt}

\makeatletter 
\newlength{\figrulesep} 
\setlength{\figrulesep}{0.5\textfloatsep} 

\newcommand{\topfigrule}{\vspace*{-1pt}%
\noindent{\color{cream}\rule[-\figrulesep]{\columnwidth}{1.5pt}} }

\newcommand{\botfigrule}{\vspace*{-2pt}%
\noindent{\color{cream}\rule[\figrulesep]{\columnwidth}{1.5pt}} }

\newcommand{\dblfigrule}{\vspace*{-1pt}%
\noindent{\color{cream}\rule[-\figrulesep]{\textwidth}{1.5pt}} }

\makeatother

\twocolumn[
  \begin{@twocolumnfalse}
{\includegraphics[height=10pt]{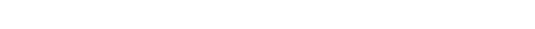}\hfill\raisebox{0pt}[0pt][0pt]{\includegraphics[height=55pt]{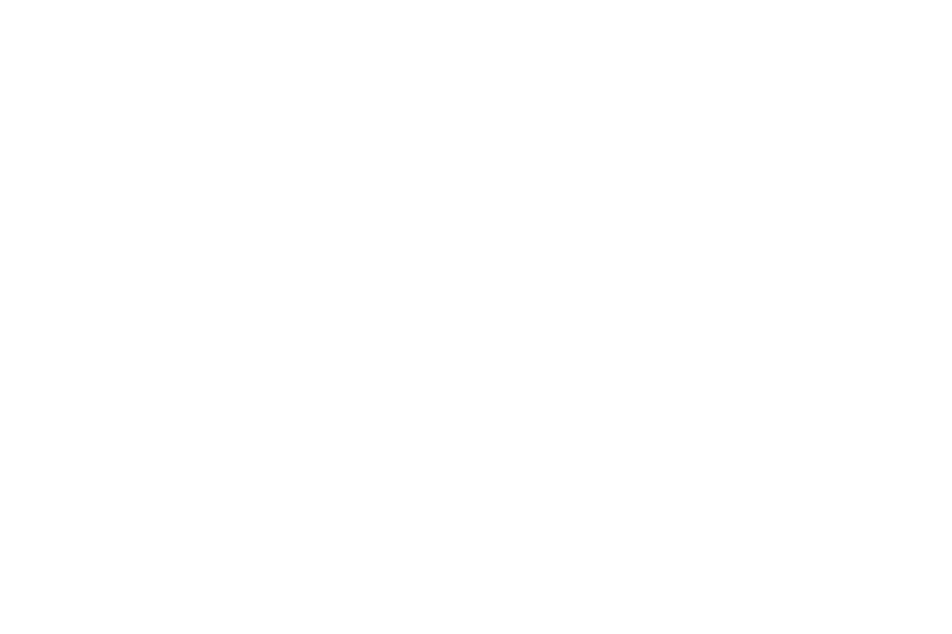}}\\[1ex]
\includegraphics[width=18.5cm]{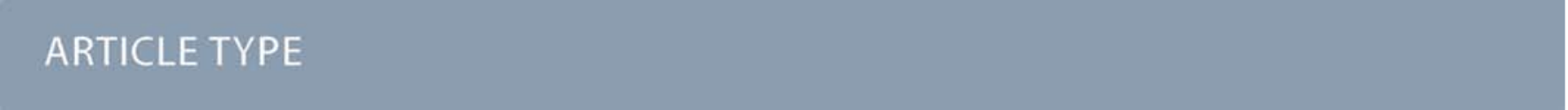}}\par
\vspace{1em}
\sffamily
\begin{tabular}{m{4.5cm} p{13.5cm} }

\includegraphics{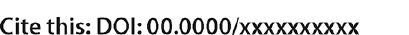} & \noindent\LARGE{\textbf{Incorporating Ion-Specific van der Waals and Soft Repulsive Interactions in the Poisson-Boltzmann Theory of Electrical Double Layers}} \\
\vspace{0.3cm} & \vspace{0.3cm} \\

 & \noindent\large{Aniruddha Seal,$^{\text{a,b},\ddag}$ Utkarsh Tiwari,{$^{\text{c,b,}\ddag}$}, Ankur Gupta$^{\text{d}}$ and Ananth Govind Rajan$^{\text{b}*}$} \\

\includegraphics{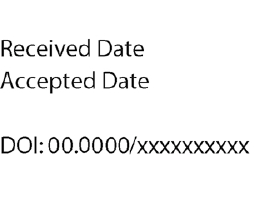} & \noindent\normalsize{Electrical double layers (EDLs) arise when an electrolyte is in contact with a charged surface, and are encountered in several application areas including batteries, supercapacitors, electrocatalytic reactors, and colloids. Over the last century, the development of Poisson-Boltzmann (PB) models and their modified versions have provided significant physical insight into the structure and dynamics of the EDL. Incorporation of physics such as finite-ion-size effects, dielectric decrement, valence asymmetry, and ion-ion correlations have made such models increasingly accurate, when compared to more computationally expensive approaches such as molecular simulations and classical density functional theory. However, a prominent knowledge gap has been the exclusion of van der Waals (vdW) and soft repulsive interactions in modified PB models. Although more short-ranged as compared to electrostatic interactions, we show here that vdW interactions can play an important role in determining the structure of the EDL via the formation of a Stern layer and in modulating the differential capacitance of an electrode in solution. To this end, we incorporate ion-ion and wall-ion vdW attraction and soft repulsion via a 12-6 Lennard-Jones (LJ) potential, resulting in a modified PB-LJ approach. The wall-ion LJ interactions were found to have a significant effect on the electrical potential and concentration profiles, especially close to the wall. However, ion-ion LJ interactions do not affect the EDL structure at low bulk ion concentrations (< 1 M). We also derive dimensionless numbers to quantify the impact of ion-ion and wall-ion LJ interactions on the EDL. Furthermore, in the pursuit of capturing ion-specific effects, we apply our model by considering various ions such as Na\textsuperscript{+}, K\textsuperscript{+}, Mg\textsuperscript{2+}, Cl\textsuperscript{-}, and SO\textsubscript{4}\textsuperscript{2-}. We observe how varying parameters such as the electrolyte concentration and electrode potential affect the structure of the EDL due to the competition between ion-specific LJ and electrostatic interactions. Lastly, we show that the inclusion of vdW and soft repulsion interactions as well as hydration effects lead to a better qualitative agreement of the PB models with experimental double-layer differential capacitance data. Overall, the modified PB-LJ approach presented herein will lead to more accurate theoretical descriptions of EDLs in various application areas.} 

\end{tabular}

 \end{@twocolumnfalse} \vspace{0.6cm}

  ]

\renewcommand*\rmdefault{bch}\normalfont\upshape
\rmfamily
\section*{}
\vspace{-1cm}


\footnotetext{{$^{\text{a}}$~School of Chemical Sciences, National Institute of Science Education and Research Bhubaneswar, Homi Bhabha National Institute, Khurda, Odisha 752050, India}}
\footnotetext{{$^{\text{b}}$~Department of Chemical Engineering, Indian Institute of Science, Bengaluru, Karnataka 560012, India}}
\footnotetext{{$^{\text{c}}$~Department of Chemical Engineering, Birla Institute of Technology and Science Pilani, K K Birla Goa Campus, Zuarinagar, Goa 403726}}
\footnotetext{{$^{\text{d}}$~Department of Chemical and Biological Engineering, University of Colorado Boulder, Boulder, Colorado 80309, United States}}
\footnotetext{Corresponding Email: ananthgr@iisc.ac.in}
\footnotetext{\dag~Electronic Supplementary Information available:~See DOI: 10.1039/cXCP00000x/}
\footnotetext{\ddag~Both authors (A.S. and U.T.) contributed equally to this work}


\section{Introduction}
The accumulation of excess counter-ions on charged surfaces, when placed in the vicinity of electrolytes, gives rise to an electrical double layer (EDL).\cite{Grahame1947} The EDL phenomenon lies at the heart of several applications such as supercapacitors,\cite{simon2010materials,wang2012physical,jarvey2022ion,gupta2020charging} porous electrodes,\cite{henrique2022charging} electrocatalytic reactors,\cite{govind2020we} membranes for water desalination,\cite{biesheuvel2010nonlinear} and nanofluidic transport devices.\cite{sparreboom2009principles,pennathur2005electrokinetic,henrique2022impact} It also has important implications in determining colloidal stability,\cite{russel1991colloidal,chu2006nonlinear,lin2017understanding} cellular integrity,\cite{wennerstrom2020colloidal} the structure of ionic liquids,\cite{avni2020charge,gavish2018solvent,gebbie2013ionic} pH effects close to electrodes,\cite{chamberlayne2020effects} and nanoconfined transport phenomena.\cite{faucher2019critical} Over the past century, the Gouy-Chapman (GC) theory\cite{gouy1910constitution,chapman1913li} has been widely employed to describe the build-up of counter-ions near charged surfaces and to compute the capacitance of EDLs.\cite{doi:10.1021/acs.chemrev.2c00097} Despite its simplistic nature, the GC theory is still in use owing to its convenient physically motivated formulation.\cite{stern1924theorie,bikerman1942xxxix,kornyshev2007double,kilic2007steric,ben2009beyond,bazant2011double} In this regard, several improvements to the classical GC theory have been suggested to make more realistic predictions keeping the mean-field framework intact. 

Finite ion-size effects have been accounted for by assuming a hard-sphere model for the ions. Efforts by Kornyshev and Kilic et al. assumed equal cation and anion diameters.\cite{kornyshev2007double,kilic2007steric,kilic2007steric2} Subsequently, Han et al. derived expressions for the counter-ion surface charge density and capacitance for valence-symmetric electrolytes with unequal cation and anion diameters, which Gupta and Stone extended for valence-asymmetric electrolytes.\cite{han2014mean,gupta2018electrical} These studies examined the EDL in an ion-agnostic manner. However, spectroscopic measurements and potentiometric pH titration experiments have shown that surface charge densities, as well as electrical potentials, depend on the identity of the ions in the electrolyte. The experimental observations correlate with Kirkwood's prediction of the deviations of the Poisson-Boltzmann (PB) equation from classical GC theory upon incorporating vdW interactions.\cite{kirkwood1934theory} In work by Lue and coworkers, the classical PB equations have been modified to include non-electrostatic interactions due to the size and polarizability of each ion to account for ion-specificity.\cite{alijo2012double,lue1999incorporation} Electrostatic correlations have been incorporated via several approaches including a fourth-order PB approach and an approach involving screening of the excluded-volume electrostatic interactions.\cite{bazant2011double,storey2012effects,bazant2009towards,gupta2020thermodynamics,gupta2020ionic} In addition, phenomena such as dielectric decrement\cite{bikerman1942xxxix,ben2009beyond,biesheuvel2005volume,hatlo2012electric,nakayama2015differential,ben2009ions,ben2011dielectric,gupta2018electrical} and hydration of ions\cite{caetano2016role,burak2000hydration,ruckenstein2002coupling,alfarano2021stripping,le2020molecular,doblhoff2021modeling,misra2019theory} have also been included to make the models more accurate. May and coworkers implemented ion-specificity in terms of varying the hydration structure of ions.\cite{caetano2016role,caetano2017differential,brown2015emergence} Advanced modeling using electronic density functional theory and ab-initio molecular dynamics simulations have also been used to understand EDL capacitance.\cite{le2020molecular,doblhoff2021modeling,ojha2022double,huang2023density}

However, so far, studies investigating the effect of dispersive van der Waals (vdW) interactions and soft repulsion between the ions and the surface, as well as between the ions themselves, on the structure of the EDL are few. Classical density functional theory employed to model EDLs\cite{jiang2014contact,wang2021demystifying,wu2011classical} has considered hard-sphere repulsions until recently when Faramarzi and Maghari incorporated dispersion interactions in the framework.\cite{FARAMARZI2017325} Although efforts in this direction have been made from the perspective of classical density functional theories, a simplified PB description involving vdW and soft repulsive interactions has not been attempted. In this regard, the development of newer theories to better describe the EDL structure has been recognized as a knowledge gap in the field.\cite{verma2022theoretical} In this work, we tackle this knowledge gap in the literature by incorporating vdW and soft repulsive interactions into the PB framework. We first extend Gupta and Stone’s model\cite{gupta2018electrical} for valence asymmetric electrolytes to include specific ion effects via a two-part vdW plus soft repulsive interaction. The two types of vdW and soft repulsive interactions we consider are between the electrolyte ions (ion-ion interaction) and between the electrode walls and the electrolyte ions (wall-ion interaction), which we systematically include in our model. To this end, we use Lennard-Jones (LJ) parameters for monoatomic ions, e.g., Na\textsuperscript{+}, K\textsuperscript{+} and Mg\textsuperscript{2+}from the study by Zeron et al.,\cite{doi:10.1063/1.5121392} and also derive optimal values for the sulfate LJ parameters, which is a polyatomic ion, using the parameters for S and O atoms.

We find that ion-ion LJ interactions only affect the EDL structure significantly only at very high ionic concentrations. On the other hand, we show that wall-ion LJ interactions play a significant role in determining the ionic concentrations near the wall, even at low ionic concentrations. We also show that the consideration of appropriate entropy expressions in the modified PB framework, depending on the relative size of the anions and the cations, is important to obtain accurate results regarding the EDL structure. Furthermore, we include the effect of hydration on the model in an approximate manner by considering realistic values for the ionic radii\cite{israelachvili2011intermolecular,doi:10.1021/cr00090a003} and examine the resultant potential and the ionic concentration profiles. We further compute the differential capacitance, thus providing an experimentally accessible quantity that reflects the properties of the EDL, to test the validity of our model. We conclude that the consideration of hydrated ion diameters and the inclusion of wall-ion LJ interactions captures the concentration and potential-dependent Stern layer thickness and predicts the double-hump nature of the differential capacitance curve in qualitative agreement with experiments.

\begin{table*}[htb!]
\centering
\caption{Diameters and LJ interaction parameters of the electrolyte ions and the wall atoms. The ion diameters were extracted from Israelachvili\cite{israelachvili2011intermolecular} and Nightingale\cite{Nightingale1959} and the inter-ionic LJ interaction parameters of the constituent ions are extracted from Zeron et al.work\cite{zeron2019force} and Xantheas et al.\cite{xantheas1996critical} The LJ parameters for graphene were taken from Cheng and Steele\cite{Cheng1990a} and that for silver from Rappe et al.\cite{rappe1992uff}}
\begin{tabular}{|c|c|c|c|c|}
\hline
\textbf{Ion or Atom} &
  \textbf{\begin{tabular}[c]{@{}c@{}}Diameter\\ (nm)\end{tabular}} &
  \textbf{\begin{tabular}[c]{@{}c@{}} Hydration Diameter\\ (nm)\end{tabular}} &
  \textbf{\begin{tabular}[c]{@{}c@{}}LJ Parameter\\ $\sigma$ (nm)\end{tabular}} &
  \textbf{\begin{tabular}[c]{@{}c@{}}LJ Parameter\\ $\epsilon$ (kJ/mol)\end{tabular}} \\ \hline
Na$^{+}$        &   0.19& 0.72&  0.221  & 1.472  \\ \hline
K$^{+}$         &   0.266& 0.66& 0.230&   1.985\\ \hline
Mg$^{2+}$       &   0.130& 0.86& 0.116&   3.651\\ \hline
Cl$^{-}$        &   0.362& 0.66&  0.469&   0.076\\ \hline
F$^{-}$         &   0.272& 0.70&  0.335&   0.418\\ \hline
SO$_{4}$$^{2-}$ &   0.580&  0.76& 0.508&    2.894\\ \hline
C               &   0.142& -&  0.340&   0.233\\
\hline
Ag               &   0.320& -&  0.263&   19.079\\
\hline
\end{tabular}
\end{table*}

\section{Model development}

\subsection{Description of the physical system}
\begin{figure}[htb!]
    \centering
    \includegraphics[width=0.9\linewidth]{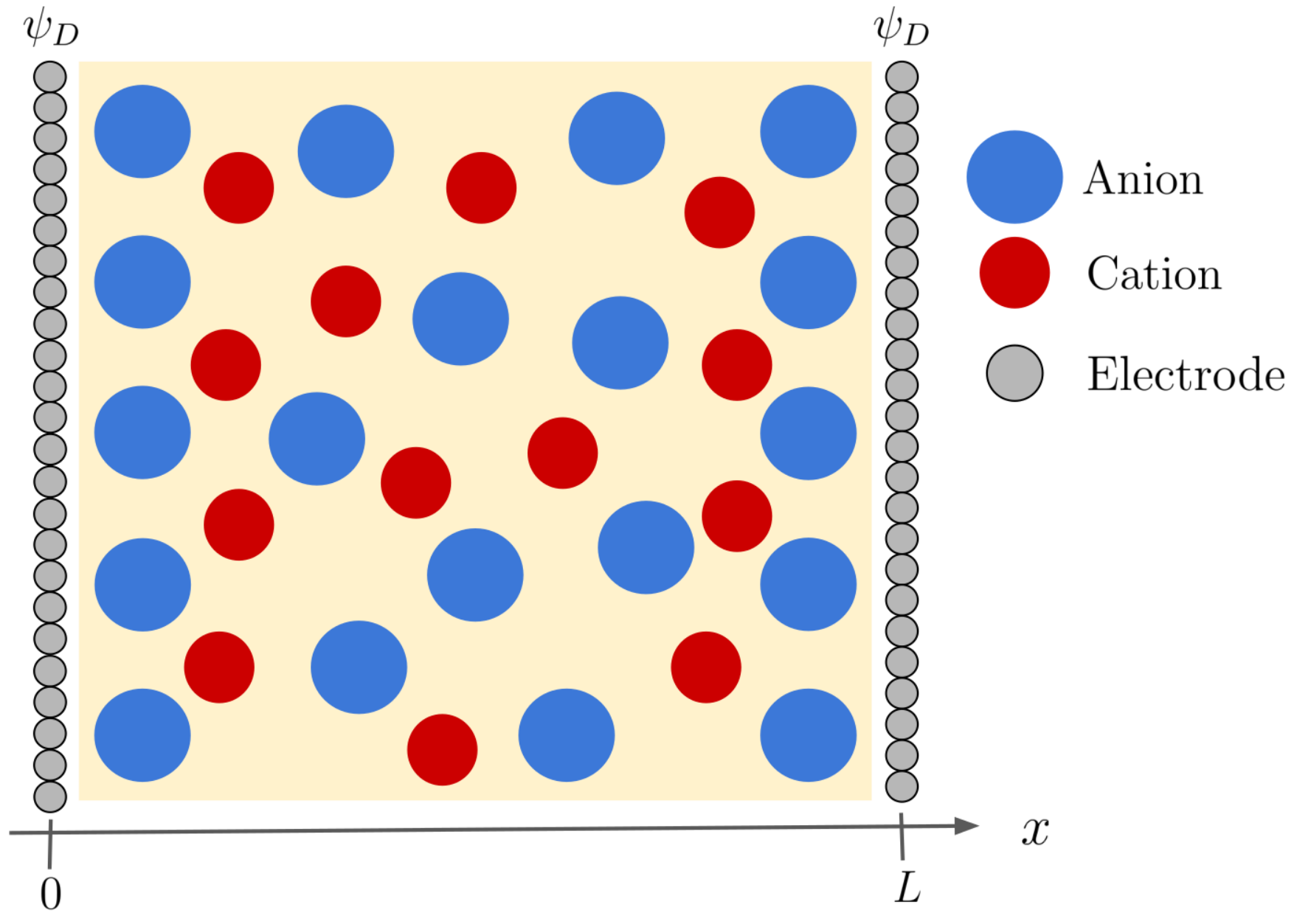}
    \caption{One-dimensional model showing electrolyte present in equilibrium between two charged surfaces separated by a distance $L$. The same potential $\Psi_D$ is applied at each wall. A Stern layer of counter-ions can be seen close to the electrodes, followed by the region of diffuse charge. The Stern layer and the region of diffuse charge together form the double layer.}
    \label{fig:EDL_schematic}
\end{figure}

We consider an electrolyte composed of one cation type and one anion type present in equilibrium between two charged surfaces in a 1D Cartesian setup shown in Figure \ref{fig:EDL_schematic}. As is well known, electrostatic attractions drive the counter-ions to migrate towards the charged surface, and these interactions compete with entropic effects, which tend to force the counter-ions away from the surface and each other, to increase their configurational entropy. The extent of the region of excess charge (which includes both the Stern layer and the diffuse charge layer), known as the EDL, is quantified by the Debye length ($\lambda_{D}$), which is given as 
\begin{equation}
    \lambda_{D} = \sqrt{\frac{\kappa_{0}\kappa_{s}k_{B}T}{e^{2}\sum_{i} z_{i}^{2}c_{0i}}}
\end{equation}
where $\kappa_{0}$ is the electrical permittivity of free space, $\kappa_{s}$ is the dielectric constant of the medium, $k_{B}$ is the Boltzmann constant, $T$ is the absolute temperature, $z_{i}$ is the valence of the $i^{th}$ ion type, $c_{0i}$ is the bulk concentration of the $i^{th}$ ion type, and $e$ is the magnitude of the charge on an electron.

Since the electrolytes considered in our study have only one type of anion and cation, $i=-$ for anions and $i=+$ for cations. The unsigned cation and anion valences of the ions constituting the electrolyte are denoted $z_{+}$ and $z_{-}$, respectively. Since the bulk region of the electrolyte is neutral, the ion concentrations therein are $c_{+} = z_{-}c_{0}$ and $c_{-} = z_{+}c_{0}$, where $c_0$ is a concentration scale. The finite ion-size effects, which have a significant effect on the ion concentration profiles, are accounted for by modeling the ions as hard spheres of an effective diameter. We denote the cation and anion diameters as $a_{+}$ and $a_{-}$, respectively. Note that electrolytes constituting ions with a range of valence and size have been used, as detailed below, to capture ion-specific effects on the EDL structure in the presence of vdW and soft reointeractions.

\subsection{Model parametrization}

We consider five types of aqueous electrolytes in our study to illustrate the effect of vdW interactions on the double layer structure -- 1:1 electrolytes (NaCl and KCl), a 2:2 electrolyte (MgSO$_4$), a 1:2 electrolyte (Na$_2$SO$_4$) and a 2:1 electrolyte (MgCl$_2$). vdW interactions between two particles (ions/atoms) $i$ and $j$ were modeled using the 12-6 LJ potential:
\begin{equation}
    U_{LJ} = -4\epsilon_{ij}\left[\left(\frac{\sigma_{ij}}{r_{ij}}\right)^6 - \left(\frac{\sigma_{ij}}{r_{ij}}\right)^{12}\right]
\end{equation}
where $r_{ij}$ represents the distance between the two particles, $\sigma_{ij}$ is the inter-particle distance at which the LJ potential is zero, and $\epsilon_{ij}$ is the well-depth of the LJ potential. Note that, as per Lorentz-Berthelot combining rules, the LJ parameters for unlike atoms can be obtained using the parameters for like atoms, as $\sigma_{ij}=\frac{\sigma_{ii} + \sigma_{jj}}{2}$ and $\epsilon_{ij}=\sqrt{\epsilon_{ii}\epsilon_{jj}}$. For simplicity, the charged walls were modeled as a single layer of graphene (or as Ag(111) in the study of differential capacitance; see below). The ion diameters were extracted from Israelachvili\cite{israelachvili2011intermolecular} and Nightingale\cite{Nightingale1959} and the inter-ionic LJ interaction parameters of the constituent ions are extracted from Vega and coworkers' study\cite{zeron2019force} (Na$^+$, K$^+$, Mg$^2+$, and Cl$^-$) and Xantheas et al.\cite{xantheas1996critical} (F$^-$). The LJ parameters for carbon atoms in graphene are taken from Cheng and Steele\cite{Cheng1990a} and that for Ag atoms from the Universal force field (UFF) developed by Rappe et al.\cite{rappe1992uff} The parameters used for all the ions and atoms considered in this study are provided in Table 1.

\begin{figure}[H]
    \centering
    \includegraphics[width=0.9\linewidth]{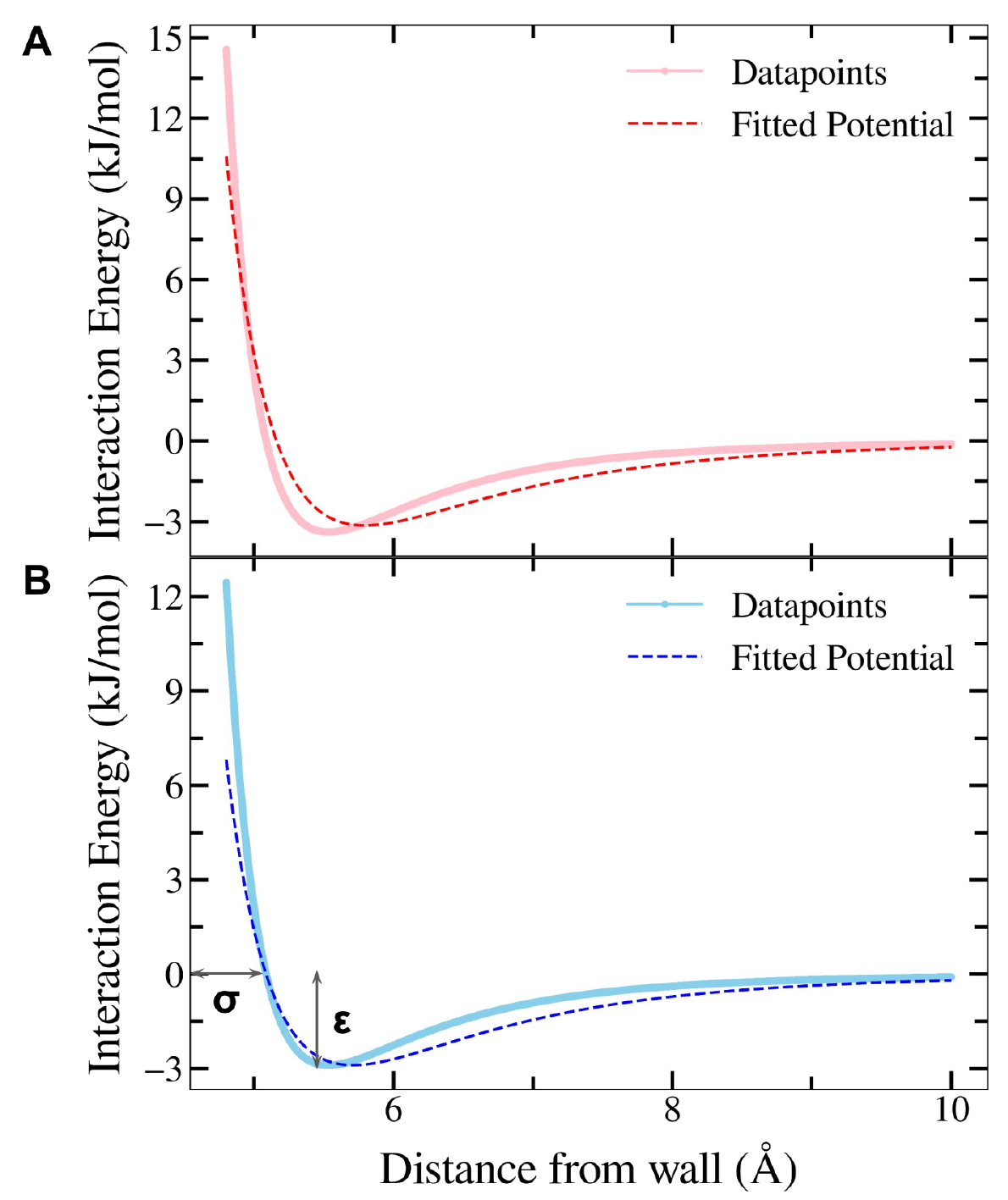}
    \caption{Comparison of the fitted and actual LJ potential for SO$_{4}$$^{2-}$ obtained using (A) least-squares fitting and (B) a physics-based approach.}
    \label{fig:fitting_LJ_SO4}
\end{figure}

We modeled the inter-ionic LJ interaction for the polyatomic SO\textsubscript{4}\textsuperscript{2-} species as the sum of all the inter-atom LJ interactions and thus determined a single set of LJ parameters. We adopted two approaches to determine a single set of LJ interaction parameters for the SO\textsubscript{4}\textsuperscript{2-} ion using the LJ parameters of the constituent S and O atoms. In the first (``best-fit'') approach, we performed a least-squares fit to obtain the parameters giving the least deviation between the combined potential and the single-centered LJ potential (Figure \ref{fig:fitting_LJ_SO4}A). We used a MATLAB code to achieve this fitting. In the second (``physics-based'') approach, we considered the distance at which the combined potential is zero as $\sigma$ and the combined well depth as $\epsilon$ (Figure \ref{fig:fitting_LJ_SO4}B). 

In both cases, two SO\textsubscript{4}\textsuperscript{2-} ions were placed next to each other at various distances from each other and the single set of LJ parameters assumed the S-S distance as the distance between the two SO\textsubscript{4}\textsuperscript{2-} ions. The respective LJ parameters are tabulated in Section S1. As can be seen in Figure \ref{fig:fitting_LJ_SO4}, the resulting potential based on the parameters we obtain from our physics-based model (Figure \ref{fig:fitting_LJ_SO4}B) can better capture the features of the attractive well in comparison to the parameters we obtain by least-square fitting over the entire length (Figure \ref{fig:fitting_LJ_SO4}A). In the rest of the paper, we use the physics-based parameters for SO$_{4}$$^{2-}$ ions in our model.

\subsection{Derivation of the effect of ion radius asymmetry in the absence of LJ interactions}
Multiple studies have accounted for \textit{entropic} finite-ion size effects in the framework of a lattice gas model, with the radius of cation being greater than that of anion.\cite{gupta2018electrical} However, as is evident from Table 1, there are several cases where it is the opposite. We therefore derive the expressions for the ion concentrations with $a_{-} > a_{+}$. A chemical potential-based thermodynamic derivation is used to compute the ionic concentration profiles. We start with the Helmholtz free energy $F$, defined per unit area of each electrode, and given by
\begin{equation}
    F = U - TS
\end{equation}
where $S$ is the entropy and $U$ is the internal energy of the system, both defined per unit area of each electrode. The internal energy can be expressed in terms of the energy stored in the electric field and electrostatic potential energy of the ions as
\begin{equation}
    U = \int_{0}^{L} \left(-\frac{\kappa_{0}\kappa_{s}}{2}{\left\lvert \frac{d\psi}{dx}\right\rvert}^{2} + z_{+}ec_{+}\psi - z_{-}ec_{-}\psi\right) dx
\end{equation}
where $\psi(x)$ is the potential at $x$ relative to the reference potential at the midpoint, $x=\frac{L}{2}$. $S$ is obtained using Boltzmann’s entropy formula, by first arranging the larger ions and then the smaller ions, and is given by
\begin{dmath}
    {-\frac{S}{k_{B}} =} \int_{0}^{L} \left( {c_{-}\ln\left({a_{-}^{3}c_{-}}\right) + \frac{1-a_{-}^{3}c_{-}}{a_{-}^{3}}\ln\left({1-a_{-}^{3}c_{-}}\right)+} \\ {c_{+}\ln\left({\frac{a_{+}^{3}c_{+}}{1-a_{-}^{3}c_{-}}}\right) + \frac{1-a_{-}^{3}c_{-}-a_{+}^{3}c_{+}}{a_{+}^{3}}\ln\left({\frac{1-a_{-}^{3}c_{-}-a_{+}^{3}c_{+}}{1-a_{-}^{3}c_{-}}}\right)} \right) dx
\end{dmath}
The chemical potentials of the constituent ions ($\mu_{\pm}$) in the electrolyte solution can be evaluated as the functional derivative of the free energy expression obtained by substituting Eqs. 4 and 5 in Eq. 3, with respect to the concentration profile:
\begin{eqnarray}
    \mu_+ = \frac{\delta F}{\delta c_+}\\
    \mu_- = \frac{\delta F}{\delta c_-}
\end{eqnarray}
Thus, we obtain
\begin{equation}
    \mu_{+} = z_{+}e\psi + k_{B}T\ln\left({\frac{a_{+}^{3}c_{+}}{1-a_{-}^{3}c_{-}-a_{+}^{3}c_{+}}}\right)
\end{equation}
\begin{equation}
    \mu_{-} = -z_{-}e\psi + k_{B}T \left[\ln\left({\frac{a_{-}^{3}c_{-}}{1-a_{-}^{3}c_{-}}}\right) - \frac{a_{-}^{3}}{a_{+}^{3}}\ln\left({\frac{1-a_{-}^{3}c_{-}-a_{+}^{3}c_{+}}{1-a_{-}^{3}c_{-}}}\right)  \right]
\end{equation}
We non-dimensionalize the ionic concentrations as $n_{+}=\frac{c_{+}}{z_{-}c_{0}}$ and $n_{-}=\frac{c_{-}}{z_{+}c_{0}}$, and the electric potential as $\Psi=\frac{e\psi}{k_{B}T}$. Further, constant chemical potential across all $x$ at equilibrium gives the following constraints
\begin{equation}
    \mu_{+}(x)=\mu_{+}(L/2) \quad \text{and} \quad \mu_{-}(x)=\mu_{-}(L/2)
\end{equation}
Using these constraints we obtain,

\begin{equation}
    n_{+} = \frac{\exp(-z_{+}\Psi)f(\Psi)}{g(\Psi)}
\end{equation}
\begin{equation}
    n_{-} = \frac{\exp(z_{-}\Psi)}{g(\Psi)}
\end{equation}
\begin{equation}
\begin{aligned}
    g(\Psi) = f(\Psi) + a_{-}^{3}z_{+}c_{0}\left(\exp(z_{-}\Psi) - f(\Psi)\right) \\+ f(\Psi)a_{+}^{3}z_{-}c_{0}\left(\exp(-z_{+}\Psi) - 1\right)
\end{aligned}
\end{equation}
\begin{equation}
    f(\Psi) = \left(1 + \frac{a_{+}^{3}z_{-}c_{0}(\exp(-z_{+}\Psi) - 1)}{1 - a_{-}^{3}z_{+}c_{0}} \right)^{\frac{a_{-}^{3}}{a_{+}^{3}}-1}
\end{equation}
Note that these equations are analogous to those presented by Gupta and Stone,\cite{gupta2018electrical} but with the $+$ and $-$ signs interchanged due to the assumption of anions being larger than cations. Now, to solve for the three unknowns: $c_{\pm}$ and $\psi$, we couple Gauss Law (Eq. 15) as the third equation with Eqs. 11 and 12 which is to be solved with the boundary conditions (Eqs. 16 and 17).
\begin{equation}
    -\frac{d^{2}\psi}{dx^{2}}= \frac{(z_{+}c_{+} - z_{-}c_{-})e}{\kappa_{0}\kappa_{S}} 
\end{equation}
\begin{equation}
    \psi(x=0) = \psi_{D}
\end{equation}
\begin{equation}
    \psi(x=L) = \psi_{D}
\end{equation}
In the following two sections, we independently incorporate the ion-ion and the wall-ion LJ contributions and derive new expressions for the ionic concentrations.

\subsection{Incorporating ion-ion LJ interactions in the PB framework}
We first derive the potential and the ion concentration expressions incorporating the ion-ion vdW and soft repulsive interactions as a pairwise 12-6 LJ potential. To this end, we consider the interaction between an infinitesimally thin section of the electrolyte located at $x$ and ions located at $x'$ and then integrate it over the entire space. The limit of the integral is chosen such that the displacement between the two sets of ions never goes below the sum of their diameters. Note that this converts the 12-6 potential to a 10-4 potential because we integrate over the lateral area of the thin section of the electrolyte located at $x$ (see Section S2). 

The ion-ion LJ interactions are then incorporated into the internal energy of the system, as three additive pieces, considering the interacting cation-cation, anion-anion, and cation-anion pairs, which are given by the expression
\begin{equation}
    \begin{aligned}
        U_{LJ}^{ij} = \frac{1}{2}\int_{0}^{L} \int_{|x-x'|\geq \frac{a_{i} + a_{j}}{2}} -4\pi\epsilon_{ij}c_{i}(x')..\\..c_{j}(x) \left[ \frac{\sigma_{ij}^{6}}{2(x-x')^{4}} - \frac{\sigma_{ij}^{12}}{5(x-x')^{10}} \right] dx dx'
    \end{aligned}
\end{equation}

where, $\left( i,j\right) = \left( +,+ \right)$ for cation-cation, $\left( i,j\right) = \left( -,- \right)$ for anion-anion, and $\left( i,j\right) = \left( +,- \right)$ for cation-anion interactions, respectively, and the interactions are defined per unit area of each electrode. $\sigma_{ij}$ and $\epsilon_{ij}$ are obtained by applying Lorentz-Berthelot combining rules on the values listed in Table 1. Note that Eq. 18 has a factor of half that corrects for incorporating the same LJ interaction twice in the expression. 

Now, the chemical potentials of the ions evaluated with the ion-ion LJ interactions can be evaluated to be:
\begin{equation}
    \mu_{+} = z_{+}e\psi + k_{B}T\ln\left({\frac{a_{+}^{3}c_{+}}{1-a_{-}^{3}c_{-}-a_{+}^{3}c_{+}}}\right) + u_{LJ}^{+}(x)
\end{equation}
\begin{equation}
\begin{aligned}
        \mu_{-} = -z_{-}e\psi + k_{B}T\Bigg[\ln \left({\frac{a_{-}^{3}c_{-}}{1-a_{-}^{3}c_{-}}}\right) \\ - \frac{a_{-}^{3}}{a_{+}^{3}}\ln\left({\frac{1-a_{-}^{3}c_{-}-a_{+}^{3}c_{+}}{1-a_{-}^{3}c_{-}}}\right)\Bigg] + u_{LJ}^{-}(x)
\end{aligned}
\end{equation}
where
\begin{equation}
\begin{aligned}
    u_{LJ}^{i}(x) = \int_{\lvert x-x'\rvert \ge a_{i}} -4\pi\epsilon_{ii}c_{i}(x') \left[ \frac{\sigma_{ii}^{6}}{2(x-x')^{4}} - \frac{\sigma_{ii}^{12}}{5(x-x')^{10}}\right] dx' \\ +\frac{1}{2} \int_{|x-x'|\geq \frac{a_{i}+a_{j}}{2}} -4\pi\epsilon_{ij}c_{j}(x') \left[ \frac{\sigma_{ij}^{6}}{2(x-x')^{4}} - \frac{\sigma_{ij}^{12}}{5(x-x')^{10}} \right] dx'
\end{aligned} 
\end{equation}
and $\left( i,j\right) = \left( +,- \right)$ for $u_{LJ}^{+}$ and $\left( i,j\right) = \left( -,+ \right)$ for $u_{LJ}^{-}$. Here, Eq. 21 is simply obtained by taking the functional derivative of Eq. 18. Applying the equilibrium conditions $\mu_{+}(x) = \mu_{+}(L/2)$ and $\mu_{-}(x) = \mu_{-}(L/2)$, we get the following expressions
\begin{equation}
\begin{aligned}
    z_{+}e\psi &+ k_{B}T \left[\ln\left({\frac{a_{+}^{3}c_{+}}{1-a_{-}^{3}c_{-}-a_{+}^{3}c_{+}}}\right) - \ln\left({\frac{a_{+}^{3}z_-c_{0}}{1-a_{-}^{3}z_+c_{0}-a_{+}^{3}z_-c_{0}}}\right)\right] \\ &+ u_{LJ}^{+}(x) - u_{LJ}^{+}(L/2) = 0
\end{aligned}
\end{equation}
\vspace{0.1cm}
and
\begin{equation}
\begin{aligned}
  -z_{-}e\psi &+ k_{B}T \left[\ln\left({\frac{a_{-}^{3}c_{-}}{1-a_{-}^{3}c_{-}}}\right) - \frac{a_{-}^{3}}{a_{+}^{3}}\ln\left({\frac{1-a_{-}^{3}c_{-}-a_{+}^{3}c_{+}}{1-a_{-}^{3}c_{-}}}\right)  \right] \\ &- k_{B}T \left[\ln\left({\frac{a_{-}^{3}z_+c_{0}}{1-a_{-}^{3}z_+c_{0}}}\right) - \frac{a_{-}^{3}}{a_{+}^{3}}\ln\left({\frac{1-a_{-}^{3}z_+c_{0}-a_{+}^{3}z_-c_{0}}{1-a_{-}^{3}z_+c_{0}}}\right)  \right]\\
  &+ u_{LJ}^{-}(x) - u_{LJ}^{-}(L/2) = 0
\end{aligned}
\end{equation}

Eqs. 15--17, 22, and 23 form  a set of nonlinear integro-differential equations that need to be solved to get the concentration profiles and $\psi$. To this end, we discretized the entire space and used the nonlinear least squares optimization (lsqnonlin) function in MATLAB, while imposing the constraints that $n_{+}\le \frac{1}{(a_{+}^3c_{0}z_{-})}$ and $n_{-}\le \frac{1}{(a_{-}^3c_{0}z_{+})}$ at each point in space. This is done to prevent the sum of the volume fractions of the anions and the cations to exceed one at any point in space. Note that we used $L=20\left(\frac{a_++a_-}{2}\right)$ in the model.

\subsection{Incorporating wall-ion LJ interactions in the PB framework}
We again employ the 12-6 form of the LJ potential to model the vdW and the soft repulsive component of the wall-ion interactions. The contribution of the wall-ion interactions to the internal energy, $U_{LJ}^w$, is calculated by integrating over the interaction of an infinitesimal section of the electrolyte located at $x$ with the two walls enclosing the electrolyte:
\begin{equation}
    U_{LJ}^{w} = \sum_{i=+,-}\int_0^L -4\pi\epsilon_{iw}\rho_{w}c_{i} \left[ \frac{\sigma_{iw}^{6}}{2x^{4}} - \frac{\sigma_{iw}^{12}}{5x^{10}} + \frac{\sigma_{iw}^{6}}{2(L-x)^{4}} - \frac{\sigma_{iw}^{12}}{5(L-x)^{10}} \right] dx
\end{equation}

where $\rho_w$ is the areal density of the wall (electrode) atoms ($\rho_w(\text{Graphene}) = 3.817 \times 10^{19} \text{ m}^{-2} \text{ and }  \rho_w(\text{Ag(111)}) = 1.383 \times 10^{19} \text{ m}^{-2})$ and the interaction parameters $\epsilon_{iw}$ and $\sigma_{iw}$ are calculated using Lorentz-Berthelot combining rules to model the wall-ion interaction. Next, we show that incorporating the wall-ion interaction term gives us an analytical solution for the ionic concentrations in terms of $\Psi$. The internal energy on incorporating the wall-ion LJ interaction term stands as

\begin{multline}
    U = \int_0^L \left(-\frac{\kappa_{0}\kappa_{s}}{2}{\left\lvert \frac{d\psi}{dx}\right\rvert}^{2} + z_{+}ec_{+}\psi - z_{-}ec_{-}\psi\right)dx - \\\sum_{i=+,-} \int_0^L \left(4\pi\epsilon_{iw}c_{i}\rho_{w}\left[ \frac{\sigma_{iw}^{6}}{2x^{4}} - \frac{\sigma_{iw}^{12}}{5x^{10}} + \frac{\sigma_{iw}^{6}}{2(L-x)^{4}} - \frac{\sigma_{iw}^{12}}{5(L-x)^{10}} \right]\right) dx
\end{multline}

The chemical potentials of the ions evaluated with the new internal energy expression are given as
\begin{equation}
    \mu_{+} = z_{+}e\psi + k_{B}T\ln\left({\frac{a_{+}^{3}c_{+}}{1-a_{-}^{3}c_{-}-a_{+}^{3}c_{+}}}\right) + u_{LJ}^{+w}
\end{equation}
\begin{equation}
    \mu_{-} = -z_{-}e\psi \\+ k_{B}T \left[\ln\left({\frac{a_{-}^{3}c_{-}}{1-a_{-}^{3}c_{-}}}\right) - \frac{a_{-}^{3}}{a_{+}^{3}}\ln\left({\frac{1-a_{-}^{3}c_{-}-a_{+}^{3}c_{+}}{1-a_{-}^{3}c_{-}}}\right)  \right] + u_{LJ}^{-w}(x)
\end{equation}
where
\begin{equation}
    u_{LJ}^{iw}(x) = - 4\pi\epsilon_{iw}\rho_{w} \left[ \frac{\sigma_{iw}^{6}}{2x^{4}} - \frac{\sigma_{iw}^{12}}{5x^{10}} + \frac{\sigma_{iw}^{6}}{2(L-x)^{4}} - \frac{\sigma_{iw}^{12}}{5(L-x)^{10}} \right]
\end{equation}
with $i = + $ for $u_{LJ}^{+w}$ and $i = -$ for $u_{LJ}^{-w}$.

Note that the reference LJ contribution, $u_{LJ}^{iw}(L/2)$, is 0 here, unlike in Eq. 23, because of the same type of wall on both sides. Future work can explore the role of asymmetry in the two walls enclosing the electrolyte. Applying the equilibrium conditions on the chemical potentials, we obtain the dimensionless ionic concentrations as: 
\begin{equation}
    n_{+} = \frac{\exp\left(-z_{+}\Psi - \frac{u_{LJ}^{+w}}{k_{B}T}\right)f(\Psi)}{g(\Psi)}
\end{equation}
\begin{equation}
    n_{-} = \frac{\exp\left(z_{-}\Psi - \frac{u_{LJ}^{-w}}{k_{B}T}\right)}{g(\Psi)}
\end{equation}
\begin{multline}
    g(\Psi) = f(\Psi) + a_{-}^{3}z_{+}c_{0}\left(\exp\left(z_{-}\Psi - \frac{u_{LJ}^{-w}}{k_{B}T}\right) - f(\Psi)\right) \\+ f(\Psi)a_{+}^{3}z_{-}c_{0}\left(\exp\left(-z_{+}\Psi - \frac{u_{LJ}^{+w}}{k_{B}T}\right) - 1\right)
\end{multline}
\begin{equation}
    f(\Psi) = \left(1 + \frac{a_{+}^{3}z_{-}c_{0}(\exp\left(-z_{+}\Psi - \frac{u_{LJ}^{+w}}{k_{B}T}\right) - 1)}{1 - a_{-}^{3}z_{+}c_{0}} \right)^{\frac{a_{-}^{3}}{a_{+}^{3}}-1}
\end{equation}

\begin{figure*}[htb!]
    \centering
    \includegraphics[width=\linewidth]{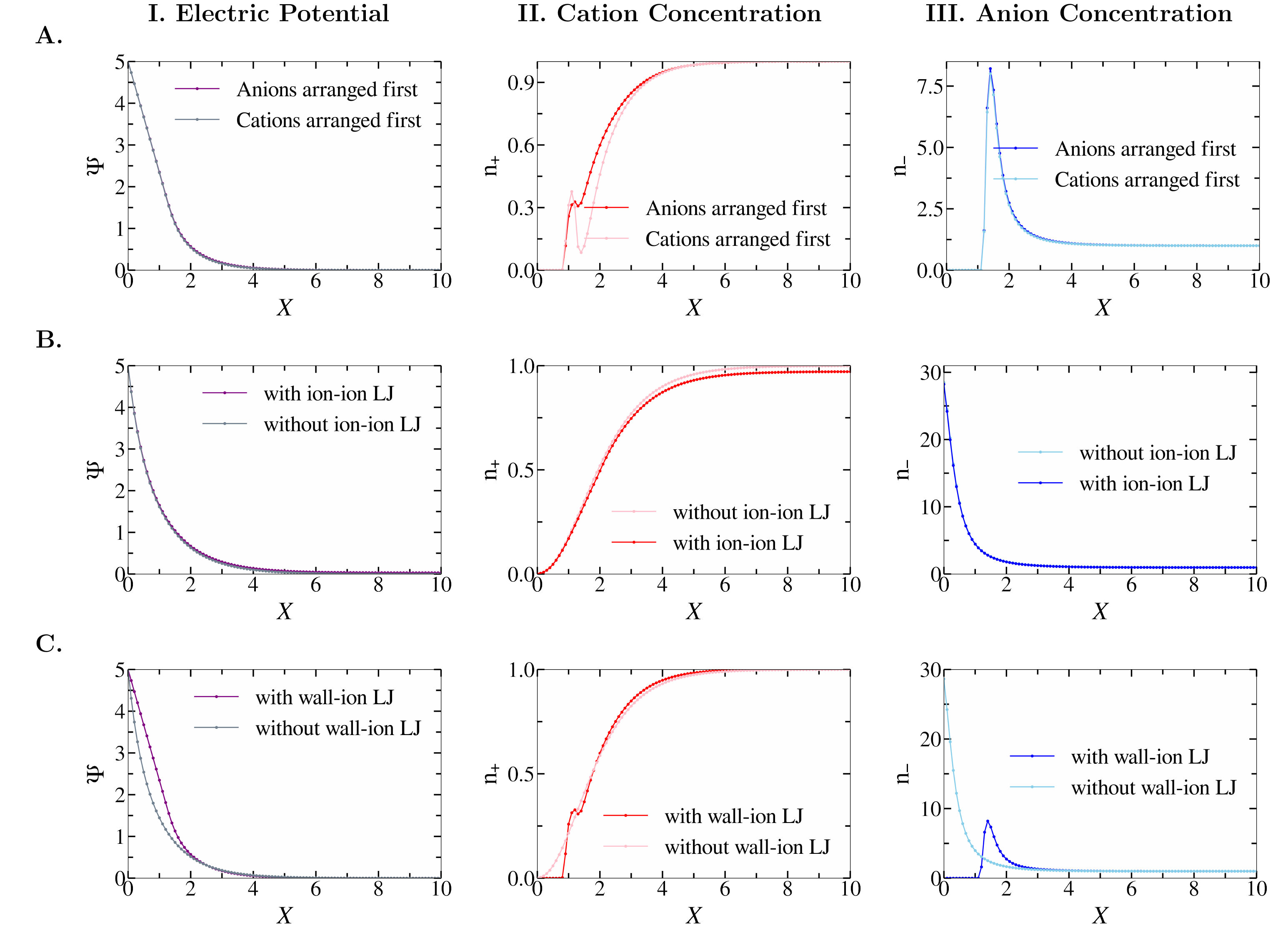}
    \caption{I. Electric potential and II, III. normalized cation and anion concentrations showing the effect of A. ion size asymmetry, B. ion-ion LJ interactions, and C.
wall-ion LJ interactions on the double layer structure for 1 M NaCl electrolyte present between two graphene electrodes with $\psi_D$ = 5. Note that $X=\frac{x}{\left(\frac{a_++a_-}{2}\right)}$.}
    \label{fig:vdW_comparison_NaCl}
\end{figure*}

\begin{figure*}[htb!]
    \centering
    \includegraphics[width=\linewidth]{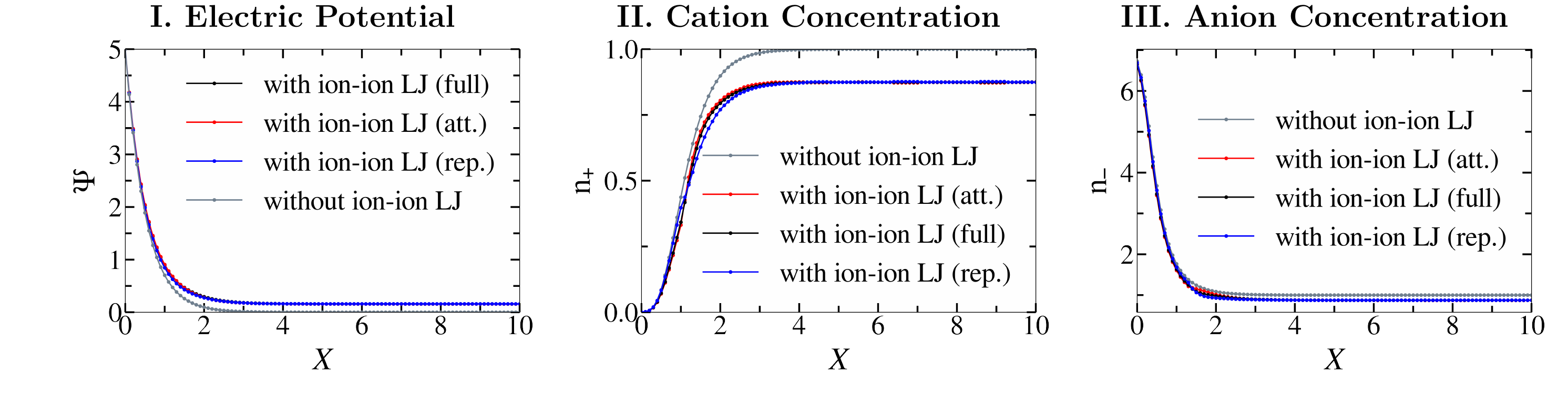}
    \caption{I. Electric potential and II, III. normalized cation and anion concentrations computed with complete, only attractive, and only repulsive ion-ion LJ interactions for 5 M NaCl electrolyte present between two graphene electrodes with $\psi_D$ = 5. Note that $X=\frac{x}{\left(\frac{a_++a_-}{2}\right)}$.}
    \label{fig:vdW_ion_broken_NaCl}
\end{figure*}

We solved the resultant set of equations (Eqs. 15--17 and 28--31) using the boundary value problem solver bvp5c in MATLAB. Note that we used $L=20\left(\frac{a_++a_-}{2}\right)$ in the model. The wall-ion interactions die out at the box center $\left(L=10\left(\frac{a_++a_-}{2}\right)\right)$, and the co- and counter-ion concentrations reach the bulk electrolyte concentration, ruling out the need for a larger system size.

\section{Results and Discussion}
\subsection{Ion diameter inequality dictates the near-wall arrangement of the smaller electrolyte ions}
The importance of considering the appropriate entropy expressions for the electrolyte ions while predicting the EDL structure is illustrated by comparing the results we obtain for 1 M NaCl using two different entropy expressions, derived assuming $a_->a_+$ and $a_+>a_-$. Considering the case of NaCl, Table 1 indicates that $a_+ = 0.19$ nm and $a_-=0.36$ nm. Thus, $a_->a_+$ and one should use the expressions for entropy, $n_+$, and $n_-$ developed in this work, which account for arranging the bigger ions (anions) in the electrolyte chamber first followed by the smaller ions (cations). However, if one uses the entropy, $n_+$, and $n_-$ expressions developed by Gupta and Stone,\cite{gupta2018electrical} i.e., by assuming $a_+>a_-$, one obtains different results. Figure \ref{fig:vdW_comparison_NaCl}A (considering wall-ion interactions) shows an erroneous layering of cations (the smaller ions) closer to the wall when we use the sign-altered entropy. This is because the smaller ions (cations) are incorrectly arranged before the larger ones, leading to more electrostatic repulsion between them, compared to the case with the corrected entropy expression, wherein the cations are arranged after the anions, allowing more of them to be accommodated within the EDL (due to favorable interactions with the oppositely charged anions), thus avoiding layering. To obtain accurate predictions, in this work, depending on each case, the appropriate expressions for the entropy, $n_+$, and $n_-$ are used, depending on whether $a_->a_+$ or $a_+>a_-$.

\subsection{Negligible effect of ion-ion LJ interactions at low ionic concentrations}
We next study the effect of the ion-ion LJ interactions on the potential and ionic concentration distributions for two positively charged walls maintained at a constant potential $\Psi_{D} = 5$ and enclosing NaCl electrolyte at 1 M concentration. The resultant plots are shown in Figure \ref{fig:vdW_comparison_NaCl}B. A negligible change in the potential profile and the anion concentration profile is observed upon incorporating the ion-ion LJ interactions. 

\begin{figure*}[htb!]
 \centering
 \includegraphics[width=\linewidth]{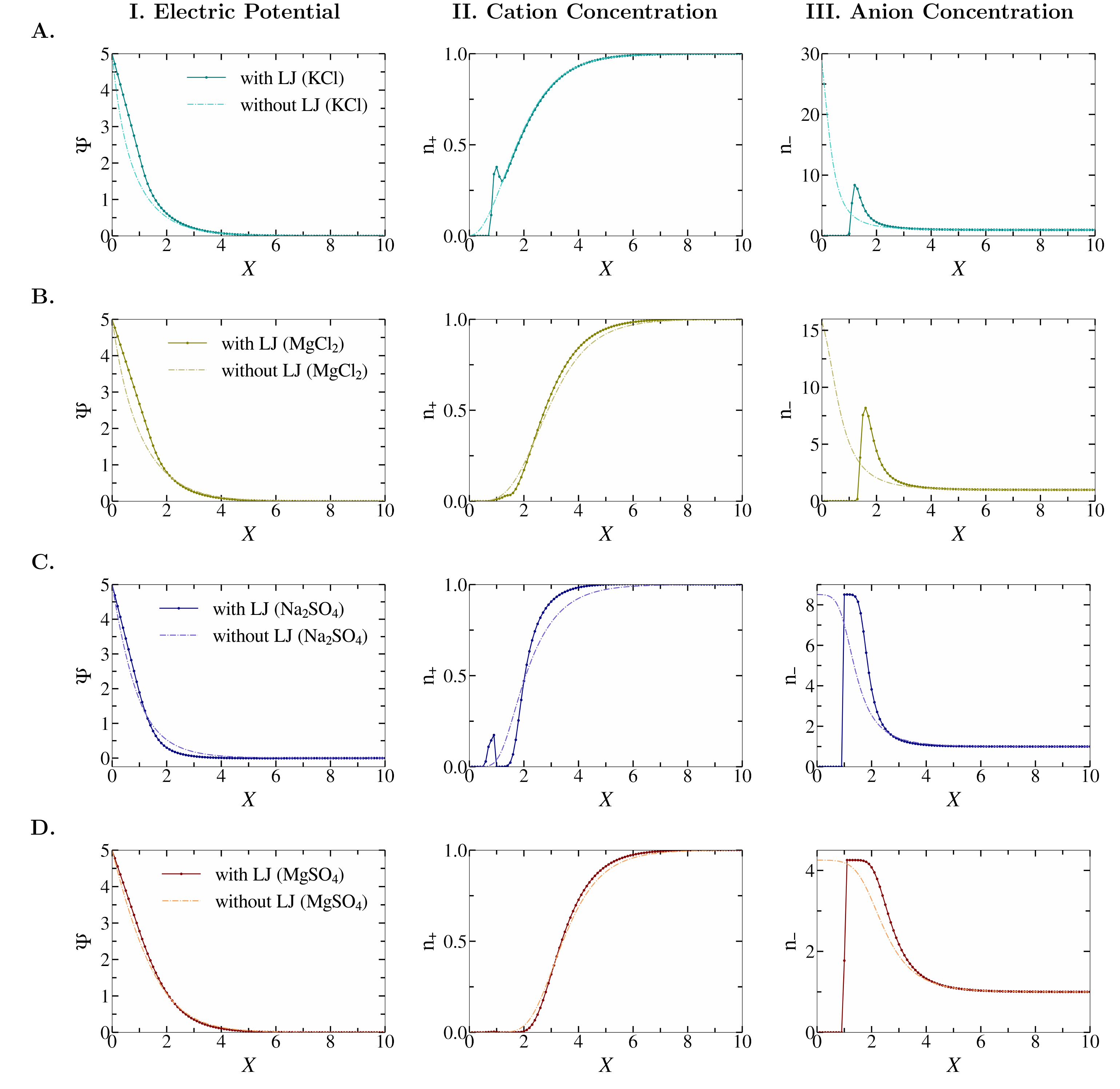}
 \caption{I. Electric Potential and II, III. normalized cation and anion concentrations showing the effect of the chemical makeup of the electrolyte on the EDL structure: A. 1:1 electrolyte (KCl), B. 2:1 electrolyte (MgCl$_{2}$), C. 1:2 electrolyte (Na$_{2}$SO$_{4}$) and D. 2:2 electrolyte (MgSO$_{4}$) present at a concentration of 1 M between the two graphene electrodes with $\Psi_{D}$ set to 5. Note that $X=\frac{x}{\left(\frac{a_++a_-}{2}\right)}$.}
\label{fig:electrolyte_type}
\end{figure*}

On the other hand, a discernible change is seen in the cation concentration profile. This is because, as per Table 1, the strength of the LJ interactions is much higher between cations (Na\textsuperscript{+}) than between anions (Cl\textsuperscript{-}). The reduction in the number of cations in the double layer upon the inclusion of ion-ion LJ interactions is due to the increased repulsion between them within the EDL, which leads to a larger chemical potential, and a lower ion concentration. The resultant increase in the cation-cation interactions leads to the drop in the cation concentration seen upon the incorporation of ion-ion LJ interactions. To quantify this observation, we calculated the ratio of the ion-ion interaction energy contributed by the LJ interactions and that contributed by electrostatic interactions as:
\begin{equation}
\eta_{\text{ion}} = \frac{\epsilon_{ii} \sigma_{ii}^{2} c_{0}\lambda_{D}}{ze\psi_{D}}
\end{equation}

\begin{figure*}[htb!]
    \centering
    \includegraphics[width=\linewidth]{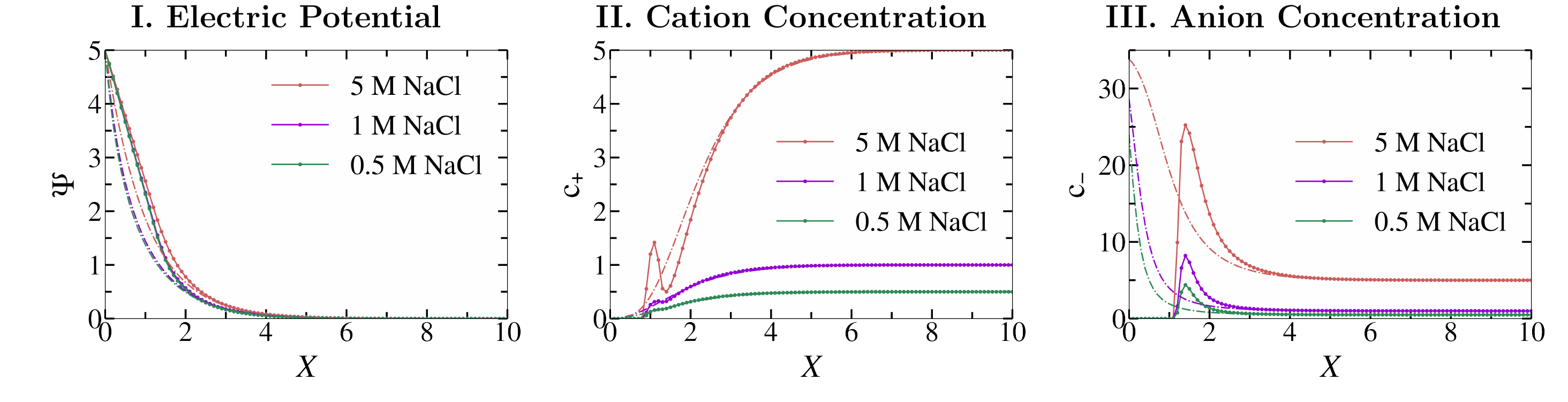}
    \caption{I. Electric Potential and II, III. cation and anion concentration profiles computed at three different electrolyte concentrations with 0.5, 1, and 5 M NaCl between the graphene electrodes with $\Psi_{D}$ set to 5. The solid lines represent the results with the wall-ion LJ terms included, and the dotted lines denote the results with the wall-ion LJ terms set to zero in the chemical potential equations. Note that $X=\frac{x}{\left(\frac{a_++a_-}{2}\right)}$.}
    \label{fig:NaCl_concentration_effect}
\end{figure*}

The rationale behind the above expression is given in Section S3. For Na\textsuperscript{+} ($i=j=+$) interactions, the ratio $\eta_{\text{ion}}$ was found to be $1.07\times 10^{-3}$, for Cl\textsuperscript{-} ($i=j=-$) interactions, this ratio was found to be $2.51\times 10^{-4}$, and for the Na\textsuperscript{+}-Cl\textsuperscript{-} ($i=+, j=-$) interactions the ratio was $5.94\times 10^{-4}$, thus explaining why ion-ion LJ interactions, although more important for cation-cation interactions, do not play a major role in determining the structure of the EDL at 1 M concentration. Furthermore, we studied the effect of electrolyte concentration on the role of ion-ion LJ interactions by investigating a bulk NaCl concentration of 5 M (Figure \ref{fig:vdW_ion_broken_NaCl}). As seen there, ion-ion LJ interactions at this concentration do affect the EDL structure significantly, with the largest effect seen on the cation concentration profile (Figure \ref{fig:vdW_ion_broken_NaCl}-II), and increasingly smaller effects on the anion concentration profile (Figure \ref{fig:vdW_ion_broken_NaCl}-III) and the potential profile (Figure \ref{fig:vdW_ion_broken_NaCl}-I). We also found that the repulsive LJ interactions play a more important role, as compared to the attractive LJ interactions. Indeed, in Figure \ref{fig:vdW_ion_broken_NaCl}-II,III one can see that the effect of the full LJ interactions is closer to that of the repulsive LJ interactions, and the absence of LJ interactions leads to behavior closer to that in the presence of attractive LJ interactions.  Nevertheless, we conclude that at reasonable bulk ion concentrations, e.g., 0.6 M as encountered in seawater, ion-ion LJ interactions are not significant and can be neglected in a PB framework.


\subsection{Counter-ion concentration profiles in dilute electrolytes are largely modulated by the wall-ion LJ interactions}
Figure \ref{fig:vdW_comparison_NaCl}C depicts the profiles of the electric potential and the ion concentrations in the EDL in the presence of wall-ion LJ interactions. One sees that the decay of the potential slows down in the presence of the wall-ion LJ interactions due to the formation of a depletion layer next to the wall. Indeed, close to the wall, the repulsive portion of the LJ potential competes with the electrostatic attraction between the counter-ions and the wall, lowering the concentration of the counter-ions near the wall and creating a depletion region, as seen in Figure \ref{fig:vdW_comparison_NaCl}C-III. In this depletion region, no ions are present, thus resulting in a slower, linear decay in the electric potential, as opposed to a faster, exponential decay in the potential, in the absence of wall-ion LJ interactions. Furthermore, the initial fall in $n_{+}$ can also be attributed to the repulsive part of the LJ potential at short distances followed by an increase from the attractive well at intermediate distances, which then gradually decays leading to the solution coinciding with the results without the LJ term. The peak of the counter-ion concentration also gets shifted away from the wall due to the repulsive part of the wall-ion LJ potential competing with the electrostatic attraction with the charged surfaces, thus leading to the formation of a Stern layer. Clearly, we see a significant effect of the wall-ion interactions on the counter-ion concentration profiles. This is also reflected in the ratio of the wall-ion interaction contributed by the LJ interactions and electrostatic interactions given as:

\begin{equation}
\eta_{\text{wall}} = \frac{\epsilon_{iw} \sigma_{iw}^{2}\rho_{w}}{ze\psi_{D}}
\end{equation}

For Na\textsuperscript{+}, $\eta_{\text{wall}}$ was found to be $1.42\times 10^{-1}$ and for Cl\textsuperscript{-}, this ratio was found to be $6.76\times 10^{-2}$, supporting the observation that wall-ion interactions significantly modulate the near-wall EDL structure, as opposed to ion-ion interactions. Indeed, our calculations reveal that $\eta_{\text{wall}}>\eta_{\text{ion}}$. Our finding holds strong implications in confined geometries where walls have a pronounced effect.\cite{perkin2012ionic}

These trends in ion concentration profiles are in qualitative agreement with the number density profiles of the electrolyte ions obtained from all-atom molecular dynamics simulations,\cite{chen2019molecular,bourg2011molecular,cats2022capacitance} although notably, our simple model does not capture the formation of \textit{multiple} layers of ions in the EDL. To capture multi-layered ionic concentration profiles, one would need to account for excluded-volume electrostatic interactions as shown recently in a PB framework by Gupta et al.,\cite{gupta2020ionic} or use a higher-order Poisson equation as proposed by Bazant et al.\cite{bazant2011double} Future work could combine these aspects with the inclusion of LJ interactions to more accurately model the EDL using a PB framework. Comparing the effect of incorporating the wall-ion LJ interaction to that of ion-ion LJ interaction, we can conclude that the ion-ion LJ interaction has a negligible effect on the double layer potential and ionic concentrations, except when considering very high bulk ionic concentrations, which may be pertinent in ionic liquids, but not in conventional electrolytes. Hence from this point, we consider only the wall-ion interaction energy term in the internal energy expression to study the physical significance of incorporating LJ interactions. 

\subsection{Effect of the chemical identity of the electrolyte}
We varied the cation and anion constituting the electrolyte to study the role of ion-specific effects in determining the EDL structure. Figure \ref{fig:electrolyte_type} depicts the EDL potential and concentration profiles for various types of electrolytes -- a 1:1 electrolyte other than NaCl (KCl), a 2:1 electrolyte (MgCl\textsubscript{2}), a 1:2 electrolyte (Na\textsubscript{2}SO\textsubscript{4}), and a 2:2 electrolyte (MgSO\textsubscript{4}) -- with $c_{0} = 1$ M and $\Psi_{D} = 5$. The initial buildup of cation concentration close to the positively charged wall for the monovalent cations (Figure \ref{fig:electrolyte_type}A,C) is due to the weaker electrostatic repulsion in play at the interface. The case of NaCl and KCl gives an ideal ground to explore the role of vdW interactions since both the cations are identical as far as electrostatics is concerned in the PB framework, and hence it will be the dispersion interactions and ion sizes that can bring in a difference. The greater magnitude of the initial peak for K$^{+}$ (Figure \ref{fig:electrolyte_type}A-II) compared to Na$^{+}$ (Figure \ref{fig:vdW_comparison_NaCl}C-II) can be attributed to the former's larger $\epsilon$ LJ parameter, leading to a greater vdW attraction and an increased concentration. Monovalent cations (Na$^{+}$, K$^{+}$) can be seen to form a distinct layer near the positively charged wall that is absent in divalent cations (Mg$^{2+}$) as seen in Figure \ref{fig:electrolyte_type}B,D-II. This is due to the larger electrostatic repulsion in the latter case overpowering the vdW attraction with the wall. 

\begin{figure*}[htb!]
    \centering
    \includegraphics[width=\linewidth]{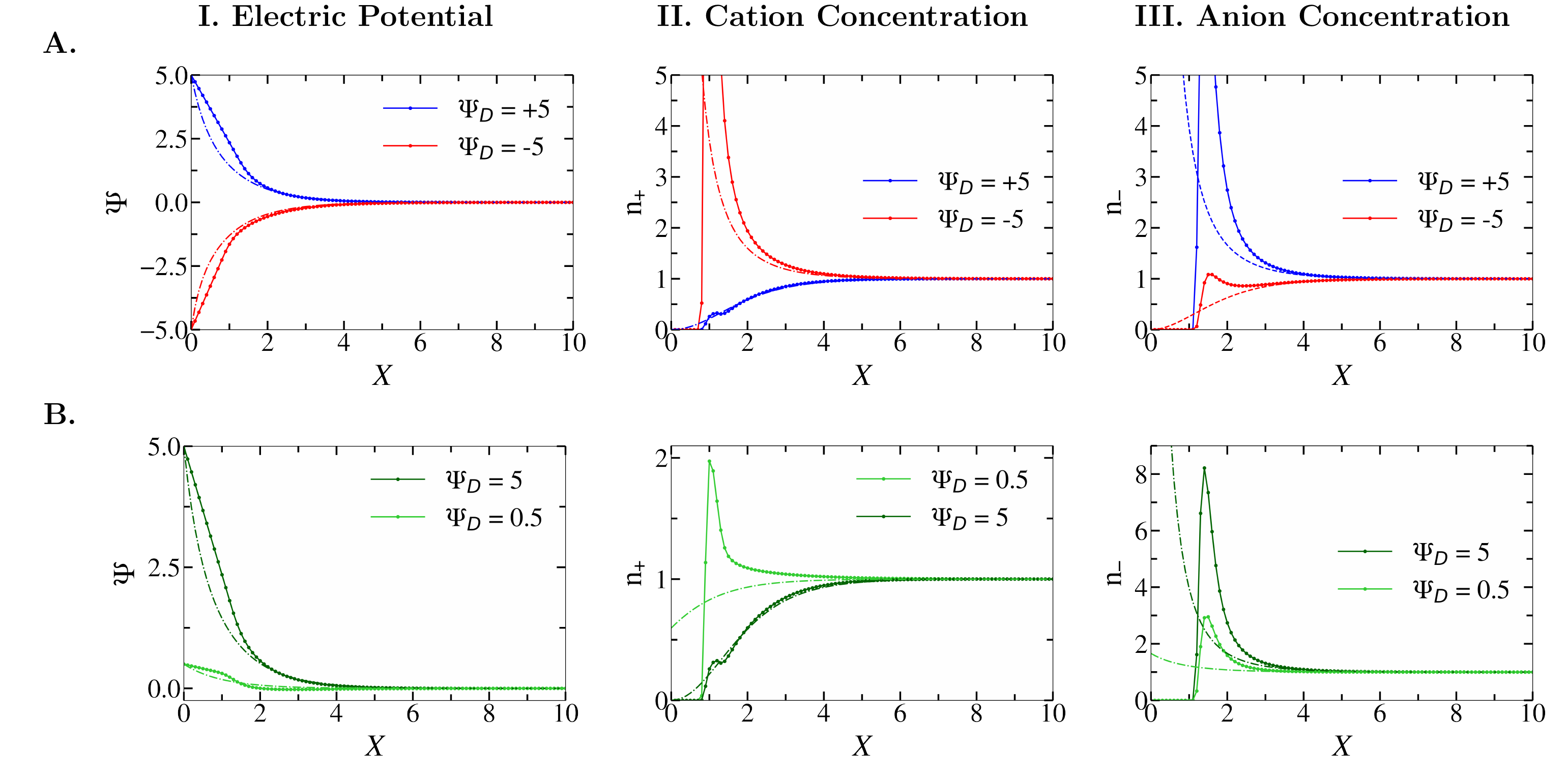}
    \caption{I. Electric potential and II, III. normalized cation and anion concentrations computed at different values of boundary potential $\Psi_{D}$ showing A. the effect of switching from positive to negative potential: $\Psi_D = +5$ to $\Psi_D = -5$ and B. the effect of changing the magnitude of wall potential: $\Psi_D = 5$ to $\Psi_D = 0.5$ for 1 M NaCl present as the electrolyte between the two graphene electrodes. The solid lines represent the results with the wall-ion Lennard-Jones terms included, and the dotted lines denote the wall-ion Lennard-Jones terms set to zero in the chemical potential equations. Note that $X=\frac{x}{\left(\frac{a_++a_-}{2}\right)}$.}
    \label{fig:effect_wall_potential}
\end{figure*}

Finally, analyzing the ion concentration results keeping the cation fixed (Mg$^{2+}$) and altering the anion (Cl$^{-}$, SO$_{4}^{2-}$), we see that due to the bulky nature of SO$_{4}^{2-}$, the concentration of cations close to the wall in the case of SO$_{4}^{2-}$ (Figure \ref{fig:electrolyte_type}D-II) is depleted upto farther away from the wall, as compared to that in the case of Cl$^{-}$ (Figure \ref{fig:electrolyte_type}B-II). One can also clearly see an anionic layer of larger width in Figure \ref{fig:electrolyte_type}C/D-III, as opposed to \ref{fig:electrolyte_type}A/B-III, due to the larger size of the sulfate ion compared to the chloride ion. Thus, our simple model is able to capture the effects of ion size/valence, soft repulsion, vdW attraction, and electrostatic interactions within a PB framework.

\subsection{Effect of the electrolyte concentration}
The effect of electrolyte concentration on the double-layer properties is studied for the case of NaCl at 0.5, 1, and 5 M. The increased screening with an increase in concentration leads to a lower Debye length, as can be seen by the slower potential decay at higher concentrations in Figure \ref{fig:NaCl_concentration_effect}-I, consistent with the findings of Gupta and Stone.\cite{gupta2018electrical} This in turn results in a wider layer of counter-ions (anions), as seen in \ref{fig:NaCl_concentration_effect}-III. Moreover, the additional screening on the inclusion of the wall-ion LJ interactions (that push the ions away from the wall) is more pronounced at lower electrolyte concentrations. As a result, the anion concentration moves closer to the wall. The same trends are followed in the case of other electrolytes. Finney et al. have outlined similar observations while studying the EDL microstructure as a function of NaCl concentration on graphene electrodes with constant chemical potential molecular dynamics simulations.\cite{finney2021electrochemistry}

\subsection{Effect of the electrode potential on the EDL structure}
The sign of the boundary potential (Figure \ref{fig:effect_wall_potential}A) determines the type of charge that will move closer to the electrode and its magnitude (Figure \ref{fig:effect_wall_potential}B) determines the concentration of the counter-ions required to neutralize its effect. The electrostatic attraction between the wall with a positive potential and the anions is toned down by the repulsive LJ potential and vice versa for the case of the attractive interactions with cations while employing a negative wall potential. In Figure \ref{fig:effect_wall_potential}, we see that at $\Psi_D=5$, less cations come closer to the wall than anions at $\Psi_D=-5$, indicating an asymmetry in the behavior of anions and cations. This finding can be rationalized by examining the ratio $\frac{\epsilon_{+w}\sigma_{+w}^6}{\epsilon_{-w}\sigma_{-w}^6} = 0.77$, which indicates a lower attraction between Na$^+$ and the wall, as compared to Cl$^-$ and the wall, thus allowing the latter to attain a higher concentration closer to the wall at $\Psi_D=-5$, compared to the former at $\Psi_D=5$. Thus, our model is able to capture ion-specific effects and, in particular, the asymmetry in cations and anions interacting with positively and negatively charged electrodes, respectively.

The electric potential in the presence of wall-ion LJ interactions decreases as we move away from the wall attaining mildly negative values before it increases again reaching zero (for the case when $\Psi_D = 0.5$), giving rise to the phenomenon of overcharging due to charge reversal. Overcharging has been predicted in EDLs using Monte Carlo simulations and observed by Kubickova et al. using electrophoresis experiments.\cite{guerrero2010overcharging,kubivckova2012overcharging} The region where the potential becomes negative for $\Psi_D = 0.5$ (see Figure \ref{fig:effect_wall_potential}B-I) is accompanied by a sharp rise in the normalized cation concentration, as seen in Figure \ref{fig:effect_wall_potential}B-II. 

\begin{figure*}[htb!]
    \centering
    \includegraphics[width=\linewidth]{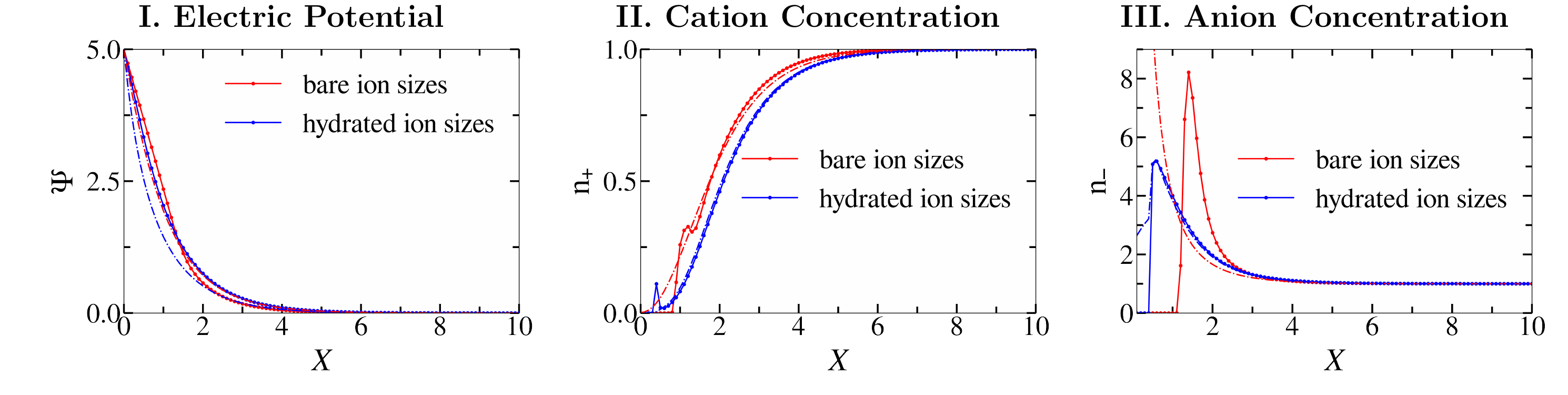}
    \caption{I. Electric potential and II, III. normalized cation and anion concentrations computed showing the effect of hydration radius for 1 M NaCl present between graphene electrodes with $\Psi_{D}$ set to 5 ($a_{\text{Na+}}^{\text{hyd.}}$/$a_{\text{Na+}}^{\text{}}$ = 3.78, $a_{\text{Cl-}}^{\text{hyd.}}$/$a_{\text{Cl-}}^{\text{}}$ = 1.82). The solid lines represent the results with the wall-ion Lennard-Jones terms included, and the dotted lines denote the wall-ion Lennard-Jones terms set to zero in the chemical potential equations. Note that $X=\frac{x}{\left(\frac{a_++a_-}{2}\right)}$.}
    \label{fig:NaCl_hydration_effect}
\end{figure*}

This region is absent in the $\Psi_D=5$ case because the large electrostatic interactions between the wall and the ions prevent the formation of a pronounced Stern layer. Qiao and coworkers have reported a similar trend upon changing the sign of the boundary potential for EDLs in organic electrolytes with molecular dynamics simulations. The authors also saw an increased counter-ion accumulation accompanied by the depletion of co-ion concentration (with the counter-ion peak moving slightly towards the electrode), with an increase in the wall potential.\cite{feng2010structure} In the future, the combination of our model with excluded-volume electrostatic interactions, as proposed by Gupta et al.\cite{gupta2020ionic} can lead to more accurate predictions of the overcharging phenomenon.

\subsection{Effect of the hydration of electrolyte ions on the EDL structure}
May and coworkers have treated ion solvation effects using hydration potentials in their mean-field formulation of EDLs.\cite{caetano2016role,caetano2017differential,brown2015emergence} Realistic potentials from explicit-water simulations have also been used in implicit-solvent Monte Carlo simulations to account for hydration in the framework of PB theory.\cite{kalcher2010ion} Recently, Misra et al. included hydration interactions in their theoretical description of the forces operative between charged surfaces enclosing multivalent elecrolytes.\cite{misra2019theory} In this work, we include the effect of hydration in an approximate manner by considering the hydrated radii of the ions in place of their ionic radii (Table 1). We see that the electrical potential falls slowly with an increased radius of the electrolyte ions. Figure \ref{fig:NaCl_hydration_effect} shows that the increased hydration radius of the cations pushes them away from the positively charged wall and the anions move closer to the wall to fill in the leftover space. Overall, this leads to a reduced concentration of anions and cations closer to the surface (Figure \ref{fig:NaCl_hydration_effect}-II,III), as well as to a slower decay in the electric potential (Figure \ref{fig:NaCl_hydration_effect}-I) as we move away from the electrode.

\subsection{LJ interactions and ion hydration modulate the features and magnitude of differential capacitance curves}

\begin{figure}[htb!]
    \centering
    \includegraphics[width=\linewidth]{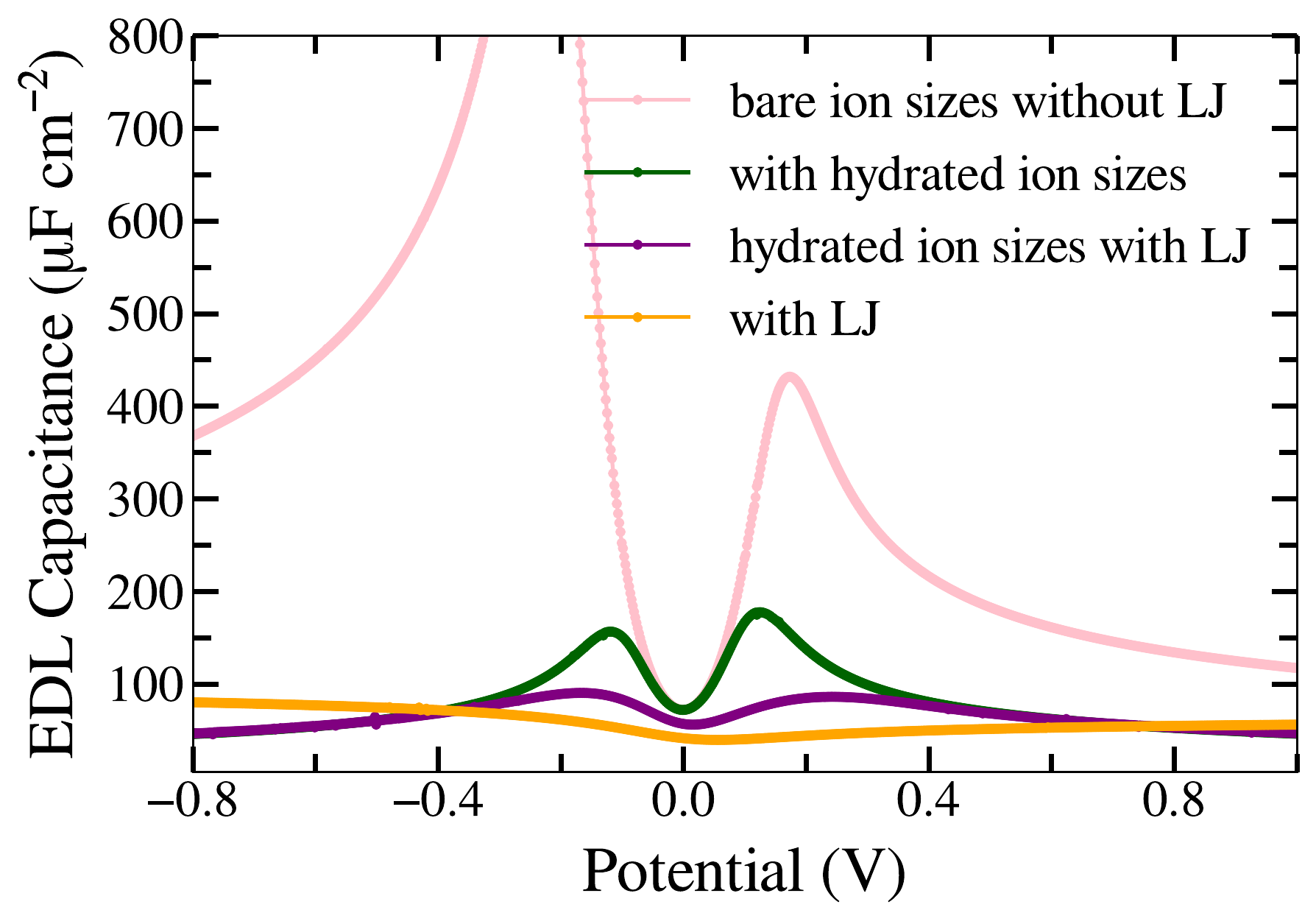}
    \caption{Differential capacitance at a graphene electrode for 0.1 M NaCl electrolyte computed by varying the wall potential from -0.8 to +0.8 V.}
    \label{fig:capacitance_effect_of_vdw_hyd}
\end{figure}

We now focus on the effect of LJ interactions and the consideration of hydrated ion diameters on the differential capacitance of an EDL ($C$), which can be computed as the change in the charge stored in the electrode per unit change in the total potential:
\begin{equation}
    C = \left| \frac{\ \text{d}q}{\text{d}\psi_{D}} \right|
\end{equation}
where the electrode surface charge density $q$ is computed as
\begin{equation}
    q = -\epsilon_{0}\epsilon_{s}\frac{\ \text{d}\psi}{\text{d}x}\bigg|_{x=0}
\end{equation}
For this set of calculations, we assumed the wall at $x=0$ to be at a potential of $\Psi_D$ and the wall at $x=L$ to be at a potential of zero. Note that modified PB theories including an explicit Stern layer calculate the derivative of the potential at a distance equal to the Stern layer thickness in Eq. (36). However, because the Stern layer naturally emerges in our PB-LJ model, the derivative of the potential used to compute the surface charge density is evaluated at $x=0$ in our model. The introduction of the wall-ion LJ interaction terms in the internal energy expression ($U_{LJ}^{+w}(x)$ and $U_{LJ}^{-w}(x)$) prevents us from obtaining an analytical expression for the double layer capacitance as reported by Gupta and Stone in the absence of LJ terms. Hence, we numerically compute the differential capacitance with our model, as shown in Figure \ref{fig:capacitance_effect_of_vdw_hyd}. 

The differential capacitance can be seen to decrease in the presence of wall-ion LJ interactions over the entire range of the electric potential we have studied for the NaCl-graphene system (Figure \ref{fig:capacitance_effect_of_vdw_hyd}) due to the formation of a potential-dependent Stern layer, whose width depends on the arrangement of ions next to the electrode at different wall potentials. We found the differential capacitance to decrease in the order: bare ion sizes without LJ interactions > hydrated ion sizes without LJ interactions > hydrated ion sizes with LJ interactions > bare ion sizes with LJ interactions. Further, note that different steric sizes of the cations and anions govern the counter-ion surface charge density at the wall leading to the asymmetry in the differential capacitance ($\text{C}(+\sigma) \neq \text{C}(-\sigma)$). One can see that the peak at negative potentials (attributable to cations) is significantly larger while considering bare ion sizes (due to the smaller size of the bare sodium ion), whereas the peak at positive potentials is marginally larger while considering hydrated ion sizes (due to the slightly smaller size of the hydrated chloride ion).

\begin{figure*}[hbt!]
\centering
\includegraphics[width=\linewidth]{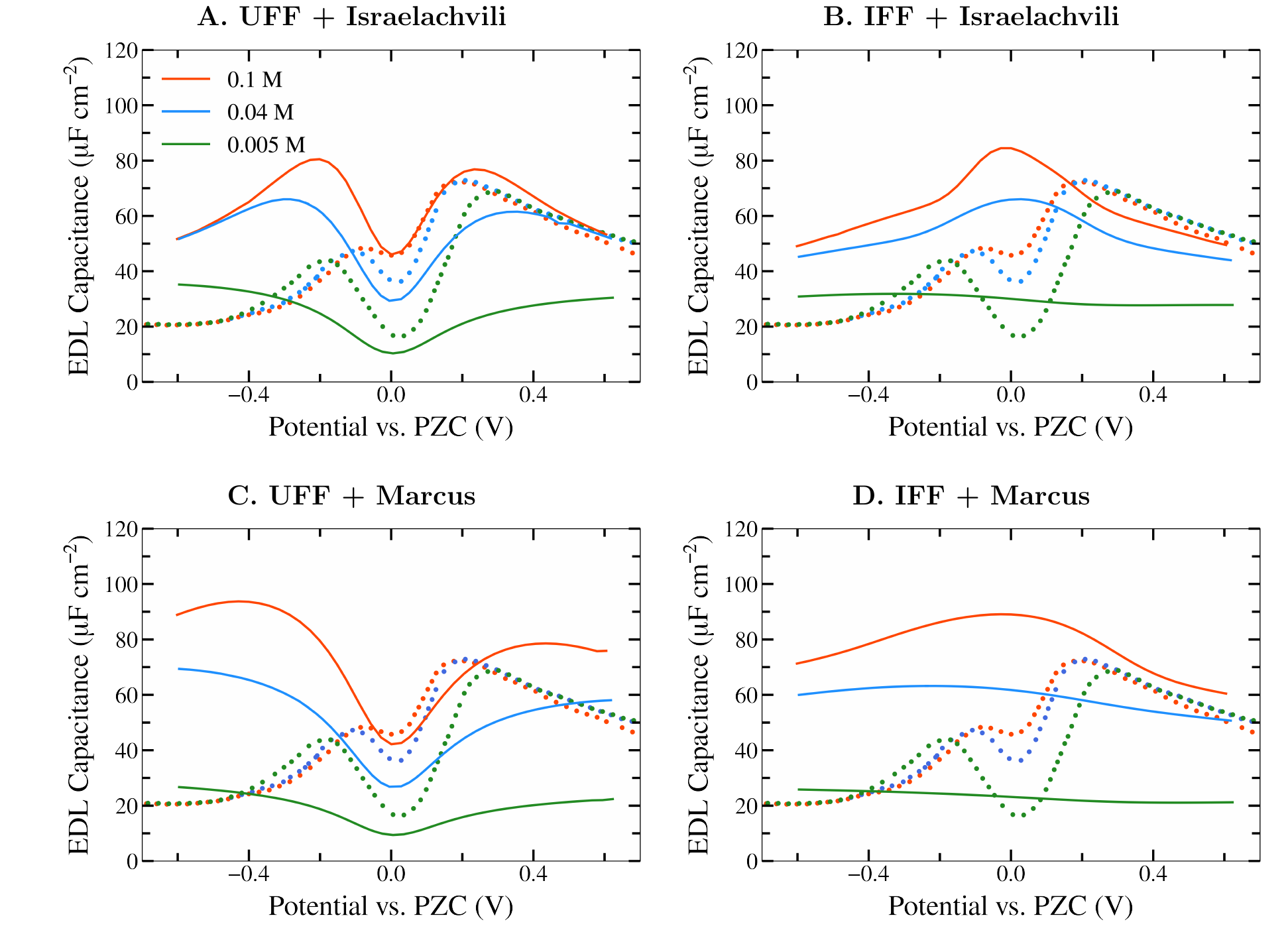}
\caption{Differential capacitance computed at the Ag(111) surface in contact with 0.1, 0.04, and 0.005 M NaF electrolyte from -0.8 to +0.9 V using our model (shown with solid lines) with parameters extracted from A. UFF for the electrode and ionic radii from Israelachvili, B. IFF for the electrode and ionic radii from Israelachvili, C. UFF for the electrode and ionic radii from Marcus and D. IFF for the electrode and ionic radii from Marcus compared with experimental data from Valette\cite{valette1989double} (shown with points). PZC stands for the potential of zero charge, as obtained from Valette\cite{valette1989double}.}
 \label{fig:capacitance_parameter_check}
\end{figure*}

We finally tested the prediction of our model against experimental differential capacitance data recorded by Valette for the Ag(111) surface in contact with NaF electrolyte at different concentrations.\cite{valette1989double} Different parameters were examined using hydration radii of electrolyte ions from Marcus\cite{marcus1988ionic} and Israelachvili,\cite{israelachvili2011intermolecular} and silver LJ interaction parameters from Heinz et al.'s interface force field (IFF)\cite{heinz2008accurate} and the UFF by Rappe et al.\cite{rappe1992uff} as shown in Figure \ref{fig:capacitance_parameter_check}. Among these various sets of parameters, the case where the ionic radii were taken from Israelachvili and silver LJ parameters obtained from the UFF performed the best (Figure \ref{fig:capacitance_parameter_check}A). The IFF overestimates the wall-ion interaction leading to the accumulation of the charges close to the wall giving a bell-shaped curve (\ref{fig:capacitance_parameter_check}B,D) instead of the reported double-hump (camel) nature of the differential capacitance curve obtained for the UFF parameters (\ref{fig:capacitance_parameter_check}A,C). Kornyshev has attributed the bell shape of differential capacitance curves to lattice saturation,\cite{kornyshev2007double} an effect that would occur when the wall-ion interactions would be very high, as in the case when the IFF parameters are used. On the other hand, the UFF causes a reasonable extent of interaction between the ions and the wall, preserving the double-hump nature of the differential capacitance curve. Moreover, the higher value for the ionic radii reported by Israelachvili slightly lowers the value of the predicted capacitance values and makes them closer to the experimental values. Thus, we find that the PB-LJ theory is able to predict a differential capacitance curve in qualitative agreement with experimental data using hydrated ionic radii and appropriate LJ parameters. 

Nevertheless, we note that the PB-LJ theory is not able to correctly predict the differential capacitance curves at low concentration values (0.005 M). This could be because the model neglects excluded-volume electrostatic interactions, which have been shown to be important even at low ionic concentrations.\cite{gupta2020ionic} Accordingly, the combined consideration of the wall-ion LJ interactions, excluded-volume ion electrostatic interactions, and hydrated ion radii could form the subject of future investigations. Moreover, the lower values of the experimental capacitance at negative potentials, as compared to the predictions from the PB-LJ theory, indicate the presence of specific adsorption effects, which could be better captured in future models.




\section{Conclusions}
In this study, we developed a modified Poisson-Boltzmann framework that incorporates the effect of vdW attractions and soft repulsions between the ions themselves and between the ions and the walls, referred to as the Poisson-Boltzmann--Lennard-Jones (PB-LJ) framework. We derived expressions for the chemical potential of the anions and the cations in the presence of such effects. The results obtained can be summarised as follows:
\begin{enumerate}
    \item The ion sizes inequality employed in the entropy expression dictates the arrangement of the smaller electrolyte ions close to the electrode and when incorrectly applied leads to an erroneous layering of the smaller ions close to the wall. 

    \item The wall-ion LJ interactions were found to greatly impact the electrical potential and concentration profiles, particularly near the wall. This observation can be rationalized via the formation of a potential-dependent Stern layer, due to the competition between soft repulsion, vdW attraction, and electrostatic interactions. In stark contrast, the ion-ion LJ interactions have little effect on the double layer structure at low bulk ion concentrations because their magnitude is negligible compared to the dominant electrostatic interactions. We proposed two dimensionless numbers to quantify the impact of the ion-ion and wall-ion interactions on the structure of the EDL.

    \item The role of ion-specific effects in determining the EDL structure was studied using electrolytes composed of different cations and anions: Na\textsuperscript{+}, K\textsuperscript{+}, Mg\textsuperscript{2+}, Cl\textsuperscript{-}, and SO\textsubscript{4}\textsuperscript{2-}. A case study varying the electrolyte concentration revealed inclusion of the wall-ion LJ interactions leads to additional screening, pronounced at lower electrolyte concentrations.
    
    \item We also studied the effect of electrode wall potential on the double-layer structure. Altering the sign of the boundary potential revealed an asymmetry in the behavior of anions and cations close to the wall, stemming from their different LJ interactions. Interestingly, with a low value of the electrode potential, a mild case of overcharging was observed, due to the competition between LJ and electrostatic interactions.

    \item Lastly, we examined the effect of hydration on EDL formation using the hydrated radii of the ions as opposed to their ionic radii. We also investigated the effect of LJ interactions and hydration on the differential capacitance of an Ag(111) electrode surface in contact with NaF electrolyte at different concentrations, showing that realistic values of the hydrated ionic radii and the wall-ion LJ parameters can provide qualitative agreement of the predicted differential capacitance curves with experimental data.
\end{enumerate}

Moving forward, our modified PB-LJ model can be made more accurate by incorporating fourth-order Poisson equation effects, excluded-volume electrostatic interactions, and dielectric decrement phenomena.\cite{gupta2018electrical,gupta2020ionic,bazant2011double,bazant2009towards} Overall, we hope that our work motivates a deeper understanding of the effect of van der Waals attraction and soft repulsion interactions on the structure and properties of EDLs.

\section*{Supplementary Information}
Lennard-Jones parameters for SO\textsubscript{4}\textsuperscript{2-}, conversion of the 12-6 LJ potential to a 10-4 LJ potential, derivation of dimensionless parameters to quantify the effect of ion-ion and wall-ion interactions on the EDL, ion radii from Marcus, and the silver LJ parameters from the interface force field. The codes used as part of this work are available via GitHub at [link to be inserted during publication].


 \section*{Conflicts of interest}
There are no conflicts to declare.

\section*{Acknowledgements}
A.G.R. acknowledges financial support from the Science and Engineering Research Board (SERB) via grant CRG/2021/002792. A.S. acknowledges the Department of Atomic Energy (DAE) for the DISHA fellowship.


\balance


\bibliography{EDL_main} 
\bibliographystyle{rsc} 

\end{document}


\maketitle
\tableofcontents

\newpage
\section{Lennard-Jones parameters of SO$_{4}$$^{2-}$}

The inter-ionic Lennard-Jones (LJ) parameters for the poly-atomic species were obtained via two different approaches -- best-fit and physics-based -- using the LJ parameters of the constituent atoms. These two sets of LJ parameters of SO\textsubscript{4}\textsuperscript{2-} are shown in Table S1.

\begin{table}[H]
\begin{tabular}{|c|c|c|}
\hline
\textbf{Fitting approach} &
  \textbf{\begin{tabular}[c]{@{}c@{}}LJ parameter\\ $\sigma$ (in nm)\end{tabular}} &
  \textbf{\begin{tabular}[c]{@{}c@{}}LJ parameter\\ $\epsilon$ (in kJ/mol)\end{tabular}} \\ \hline
\text{Physics-based} & 0.51775   & 2.7963
   \\ \hline
\multicolumn{1}{|l|}{\text{Least-square fitting}} & 0.50862 & 2.8947 \\ 
\hline
\end{tabular}
\caption{LJ interaction parameters of SO$_{4}$$^{2-}$ obtained using best-fit and physics-based approaches}
\end{table}

\section{Conversion of the 12-6 LJ potential to a 10-4 LJ potential}
We work out the integration of the pairwise 12-6 LJ potential leading to the 10-4 LJ potential used in Sections 2.4 and 2.5 in the main text for incorporating ion-ion/wall-ion LJ interactions in our PB-LJ framework. The LJ interaction between a wall (or an infinitesimally thin section of the electrolyte) and ions separated by a distance $x$ (Figure S1) is integrated over the entire wall (or the sheet of the electrolyte) giving rise to a 10-4 potential (Eq. S3).

\begin{figure*}[hbt!]
\centering
\includegraphics[width=0.8\linewidth]{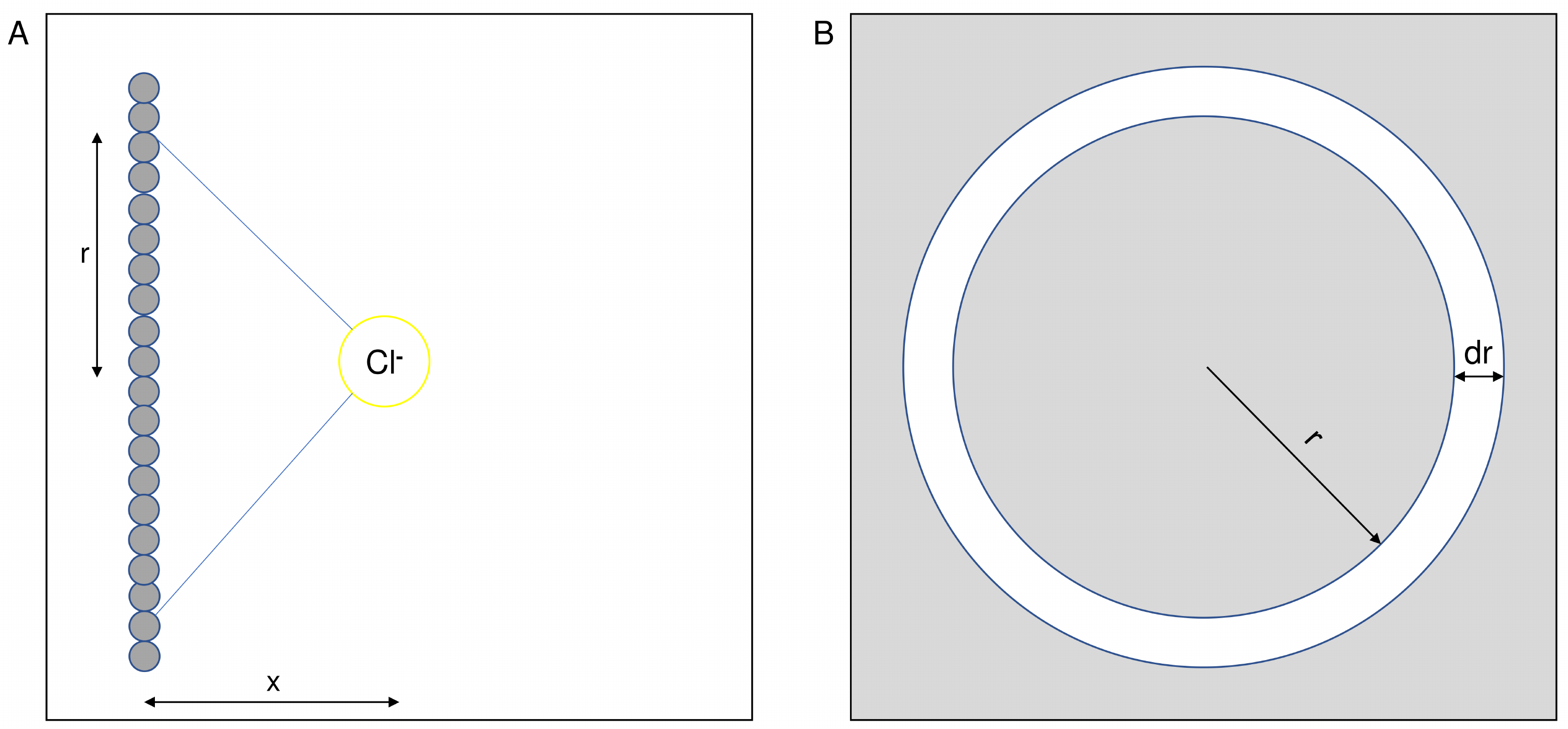}
\caption{A. Side view of an ion interacting with a wall, separated by distance $x$ and B. front view of the wall showing an infinitesimal ring of thickness $dr$.}
 \label{fig:10_4_dv}
\end{figure*}
The total LJ interaction energy is given as
\begin{equation}
    U = \int_{0}^{\infty} -4\epsilon \left( \frac{\sigma^{6}}{(\sqrt{x^{2}+r^{2})}^{6}} - \frac{\sigma^{12}}{(\sqrt{x^{2}+r^{2})}^{12}}\right)2\pi\rho_{w} rdr
\end{equation}
where $\rho_w$ is the density of the atoms in the wall. The expression can be simplified to get
\begin{equation}
    U = -8\pi\epsilon\rho_{w} \left( \sigma^{6}\int_{0}^{\infty}\frac{rdr}{(x^{2}+r^{2})^{3}} - \sigma^{12}\int_{0}^{\infty}\frac{rdr}{(x^{2}+r^{2})^{6}}\right)
\end{equation}
Finally, we obtain
\begin{equation}
    U = -4\pi\epsilon\rho_{w} \left( \frac{\sigma^{6}}{2x^{4}} -\frac{\sigma^{12}}{5x^{10}}\right)
\end{equation}

\newpage

\section{Derivation of dimensionless parameters to quantify the effect of ion-ion and wall-ion interactions on the EDL}

The ratios of the ion-ion and wall-ion interaction energies contributed by LJ interactions and the electrostatic potential energy of an ion ($\eta_{\text{ion}}$ and $\eta_{\text{wall}}$) were computed as follows. The ion-ion LJ interaction energy ($u_{\text{LJ}}^{\text{ion-ion}}$) is given by Eq. 21 in the main text. We have approximated the value of this integral by considering the electrolyte concentration scale to be the bulk electrolyte concentration ($c_{0}$), the length scale to be the Debye length ($\lambda_D$), and the ion-ion interaction scale as its maximum value, attained at a distance of $\sigma_{ii}$. Hence the ion-ion LJ interaction energy can be estimated as

\begin{equation}
\begin{aligned}
    u_{\text{LJ}}^{\text{ion-ion}} \sim - 4\pi\epsilon_{ii}c_{0} \left[ \frac{\sigma_{ii}^{6}}{2\sigma_{ii}^{4}} - \frac{\sigma_{ii}^{12}}{5\sigma_{ii}^{10}}\right]\lambda_{D} = - \frac{6\pi\epsilon_{ii}c_{0} \sigma_{ii}^{2} \lambda_{D}}{5}
\end{aligned}
\end{equation}

Following a similar procedure starting from the one-wall contribution in Eq. 28 in the main text, the wall-ion LJ interaction energy ($u_{\text{LJ}}^{\text{wall-ion}}$) can be estimated as

\begin{equation}
\begin{aligned}
    u_{\text{LJ}}^{\text{wall-ion}} \sim - 4\pi\epsilon_{iw}\rho_{w} \left[ \frac{\sigma_{iw}^{6}}{2\sigma_{iw}^{4}} - \frac{\sigma_{iw}^{12}}{5\sigma_{iw}^{10}} + \frac{\sigma_{iw}^{6}}{2L^{4}} - \frac{\sigma_{iw}^{12}}{5L^{10}}\right] \approx \frac{6\pi\epsilon_{iw}\rho_{w}\sigma_{iw}^{2}}{5} 
\end{aligned}
\end{equation}

The electrostatic interaction energy ($u_{\text{elec}}$) scale can be simply written as

\begin{equation}
    u_{\text{elec}} = ze\psi_{D}
\end{equation}

Combining Eqs. S4 and S6, $\eta_{\text{ion}}$ can be written as
\begin{equation}
\begin{aligned}
    \eta_{\text{ion}} &= \frac{u_{\text{LJ}}^{\text{ion-ion}}}{u_{\text{elec}}} \quad = \frac{6\pi\epsilon_{ii}c_{0} \sigma_{ii}^{2} \lambda_{D}}{5ze\psi_{D}}\sim \frac{\epsilon_{ii}\sigma_{ii}^{2}c_{0} \lambda_{D}}{ze\psi_{D}} \quad \text{(numerical factors removed)}
\end{aligned}    
\end{equation}

Combining Eqs. S5 and S6, $\eta_{\text{wall}}$ can be written as
\begin{equation}
\begin{aligned}
    \eta_{\text{wall}} &= \frac{u_{\text{LJ}}^{\text{wall-ion}}}{u_{\text{elec}}} \quad = \frac{6\pi\epsilon_{iw} \sigma_{iw}^{2}\rho_{w}}{5ze\psi_{D}}\sim \frac{\epsilon_{iw} \sigma_{iw}^{2}\rho_{w}}{ze\psi_{D}}  \quad \text{(numerical factors removed)}
\end{aligned}    
\end{equation}

\section{Ion radii from Marcus and the silver LJ parameters from the interface force field}

The parameters used to compute the EDL capacitance in panels B, C, and D in Figure 10 in the main text are given here. In the interface force field (IFF) developed by Heinz et al.,\cite{heinz2008accurate} the $\epsilon$ and $\sigma$ values for silver are 19.079 kJ/mol and 0.263 nm, respectively. The hydrated diameters of the electrolyte ions extracted from Marcus are given in Table S2 below.

\begin{table}[H]
\centering
\caption{Diameters of the electrolyte ions extracted from Marcus.\cite{marcus1988ionic}}
\begin{tabular}{|c|c|}
\hline
\textbf{Ion} &
  \textbf{\begin{tabular}[c]{@{}c@{}}Hydrated Diameter\\ (nm)\end{tabular}} \\ \hline
Na$^{+}$        &    0.471  \\ \hline
F$^{-}$        &   0.526\\ \hline
\end{tabular}
\end{table}

\bibliographystyle{rsc}
\bibliography{EDL_main}

